\documentclass[twocolumn]{aastex631}
\usepackage{graphicx}
\usepackage{xcolor}
\usepackage [english]{babel}
\usepackage [autostyle, english = american]{csquotes}
\usepackage{amsmath}
\usepackage{makecell}
\usepackage{booktabs}
\MakeOuterQuote{"}

\begin{document}

\title{AGN Feedback Models and AGN Demographics I:

Radio-Mode AGN in EAGLE, SIMBA and TNG100 are Inconsistent with Observations}

\author{Arjun Suresh}
\author{Michael R. Blanton}
\affil{New York University}
\author{Douglas Rennehan}
\affil{Center for Computational Astrophysics, Flatiron Institute}

\begin{abstract}
We compare predictions of how Active Galactic Nuclei (AGN) populate host galaxies at 
low redshifts to observations, finding large discrepancies between cosmological 
simulation predictions and observed patterns. Modern cosmological simulations include 
AGN feedback models tuned to reproduce the observed galaxy stellar mass function. 
However, due to a lack of real understanding of the physics of AGN feedback, these
models vary significantly across simulations. To distinguish between the 
models and potentially test the underlying physics, we carry out  independent tests 
of these models. In an earlier study, we found that $F_{\rm AGN}$---the observed 
completeness-corrected fraction of galaxies hosting radio AGN with an Eddington ratio 
$\lambda > 10^{-3}$---to be a strong function of host galaxy stellar mass ($M_\star$) 
but nearly independent of host specific star formation rates (sSFR) at fixed $M_\star$. 
In this study, we test the radio mode AGN feedback models of the EAGLE, SIMBA, and 
TNG100 simulations by comparing their predictions of $F_{\rm AGN} \left(M_\star \right)$ 
to our observational constraint. We find that none of these simulations even qualitatively 
reproduce the observed dependencies of $F_{\rm AGN}$ on $M_\star$ and sSFR. Finally,
we find that although the given TNG100 model could be modified in order to 
better reproduce the observed $F_{\rm AGN}$ trend, this modification would 
likely also change its prediction for the local stellar mass function 
and star formation rates---key observations used for calibrating the simulation 
in the first place. Our findings highlight a pressing need to revisit the AGN 
feedback prescriptions in EAGLE, SIMBA, TNG100 and other similar models.
\end{abstract}

\section{Introduction}\label{intro}

Active Galactic Nuclei (AGN) are supermassive black holes at the centers of galaxies, 
displaying highly luminous activity across the electromagnetic spectrum, caused by 
their rapid accretion of matter. Theoretical and observational studies of X-ray binaries 
and AGN yield strong evidence that accretion onto black holes occurs in two major 
modes---radiatively efficient and radiatively inefficient modes (\citealt{maccarone2003,merloni08,heckman14a,MHD03}). Whereas radiatively efficient accretion occurs at high  Eddington ratios 
($L_{\rm Bol}/ L_{\rm Edd} > 0.02$), radiatively inefficient accretion 
occurs at low Eddington ratios ($L_{\rm Bol}/ L_{\rm Edd} < 0.02$). Here 
$L_{\rm Bol}$ is the AGN bolometric luminosity and $L_{\rm Edd}$ is the 
Eddington luminosity of the black hole. Strong radio emission is seen primarily 
in low Eddington ratio AGN, which are also referred to as radio-mode AGN or 
radio AGN (\citealt{ho02}); however, we note that $\gtrsim 10\%$ of radiatively 
efficient AGN are also bright in the radio (\citealt{Kellermann_2016}). 
In this paper, we will focus primarily on radiatively inefficient AGN, comparing
their radio properties and their relationship with their host galaxies to simulation
predictions.

Radio AGN activity is frequently invoked in galaxy formation/ evolution theories to suppress star
formation rates at high stellar masses (\citealt{cronton2006, bower2006, someville&dave2015}). The claim is that radio AGN are involved in impeding the cooling of hot gas in massive galaxies, thereby suppressing their star formation rates. This theoretical picture is motivated mainly by the following observations:

\begin{itemize}
    \item Massive galaxies in nearby clusters seldom form stars at a rate implied by uninhibited cooling of hot gas in their atmospheres (\citealt{fabian1994, egami2006, edge2001, peterson2003}). 

    \item These massive galaxies are also the ones that are observed to be most likely to host radio AGN (\citealt{matthews64a, best05a, heckman14a}). Further, the energy input from the radio AGN is measured to correlate well with the cooling luminosity of the hot gas (\citealt{binney&tabor1995, fabian2012}). 

    \item Joint X-ray and radio observations of these objects also show that sometimes associated with radio AGN jets are cavities in the hot gas atmospheres (\citealt{mcnamara2000, churazov2000, dunn&fabian2006}). These cavities are interpreted as evidence for radio AGN activity interacting with the atmosphere. 
    
\end{itemize}

While the overall energetics work in the above picture, there have not been 
reliable observations of how the energy output from AGN is coupled to the gas. 
Hence, the physics of AGN feedback, and the role it plays in host galaxy 
evolution is not yet understood well. Furthermore, until recently, no accurate measurement of the relationship between radio AGN feedback and host galaxy star formation rates had been published.

In an observational precursor to this work, we estimated this relationship
in the local universe \citep{sureshblanton2024}. We used integral field 
spectroscopy from the Mapping Nearby Galaxies at APO (MaNGA; \citealt{bundy15a}) to
estimate star formation rates robustly for a broad set of galaxies with a well-understood
selection function.  Following \cite{best05a}, we matched this data to 20-cm radio
data from the NRAO VLA Sky Survey (NVSS; \citealt{condon98a}) and the Faint Images 
of the Radio Sky at Twenty-centimeters (FIRST; \citealt{becker95a}), to measure 
the radio AGN Eddington ratio distribution in the local Universe, 
and its dependence on host galaxy properties. This study, by virtue of MaNGA's 
high precision determinations of galaxy properties and a careful treatment of 
the sample selection effects, provides a more reliable determination of the 
relationship between radio AGN activity and host galaxy properties than was 
previously available. We found the following:
\begin{enumerate}
    \item Radio AGN activity is a very strong function of host galaxy $M_{\star}$, as
    previously known (\citealt{heckman14a}). 
    \item At a given $M_{\star}$, the radio AGN fraction is independent of host sSFRs over a large range of its values ($-16 \leq \log \left( {\rm sSFR}/ {\rm yr}^{-1}\right) \leq -9$). 
\end{enumerate}

The first result is well known in the literature, thanks to the work of 
\cite{best05a}, \cite{best12a}, \cite{sabater2019}, and others. However, the 
second result is new (though see related work by \citealt{janssen12a}, 
\citealt{mulcahey2022}, and Figure 14 of \citealt{heckman14a}). 

\begin{figure*}[h]
    \centering
    \includegraphics[width = 7.2in]{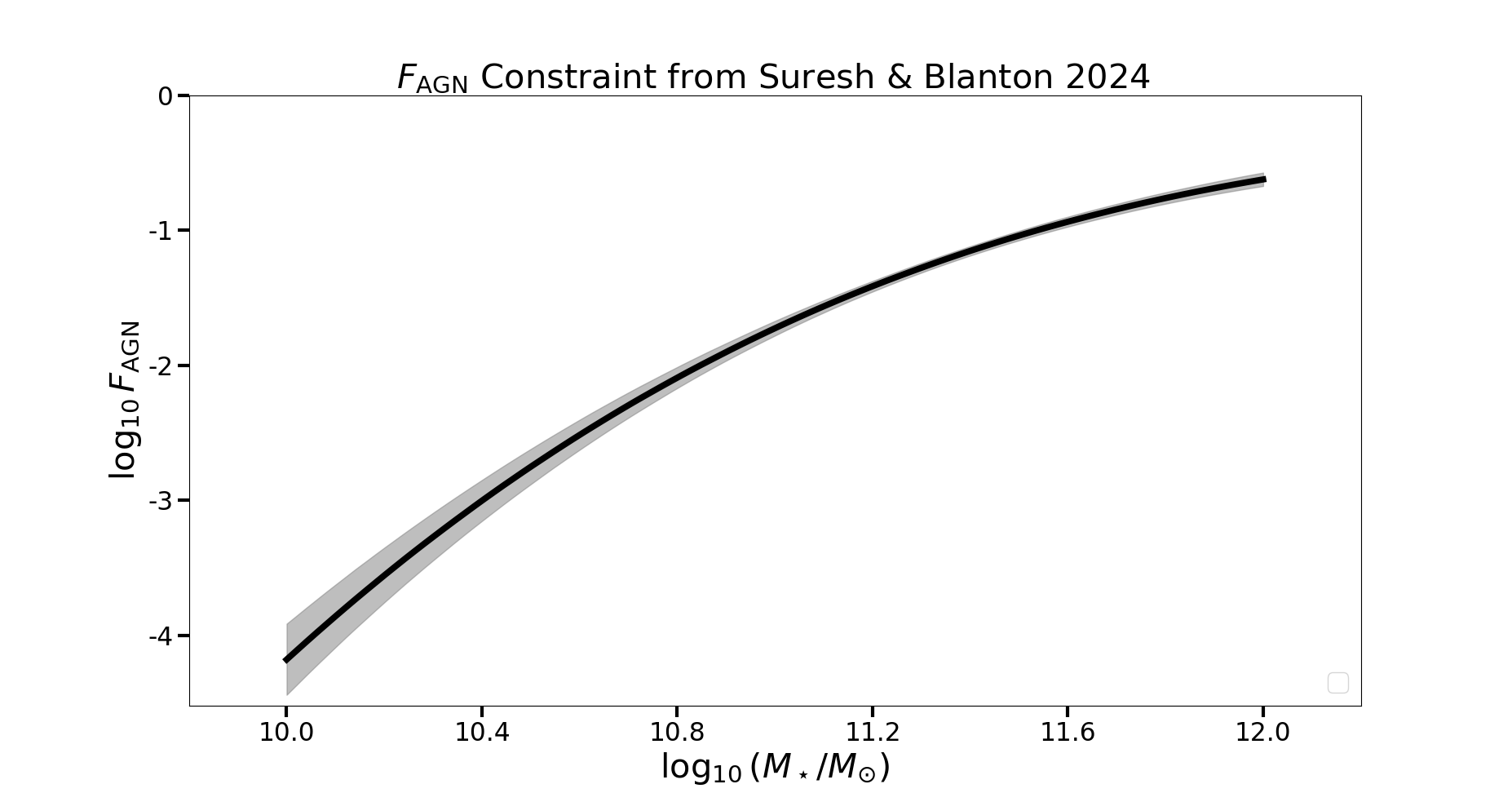}
    \caption{Fraction of galaxies ($F_{\text{AGN}}$) hosting radio AGN with Eddington ratios $\lambda > \lambda_c$ ($\lambda_c = 10^{-3}$) as a function of stellar mass ($M_*$). Here, $\lambda$ represents the mechanical power of the AGN scaled by the black hole’s Eddington luminosity. Although there are systematic uncertainties in $\lambda$'s definition and measurement, this strong dependence provides a robust test for galaxy formation and evolution models. Adapted from \cite{sureshblanton2024}, who use MaNGA optical IFU data in conjunction with NVSS and FIRST radio data.}
    \label{fig:Fagn_manga}
\end{figure*}

Figure \ref{fig:Fagn_manga} (adapted from \citealt{sureshblanton2024})
shows the fraction of galaxies $F_{\rm AGN}$ that are radio AGN with an 
Eddington ratio $\lambda > \lambda_{c}$, where $\lambda_c = 10^{-3}$, as 
a function of $M_{\star}$. We choose this value of $\lambda_c$ to roughly 
minimize the  error bars.  Here, the meaning of $\lambda$ is slightly
different from the usual definition mentioned earlier. Here, it represents 
the mechanical power (different from the usual bolometric luminosity) of the 
AGN, scaled by the black hole's Eddington luminosity. 
There are systematic uncertainties 
regarding what this $\lambda$ means, how well it is measured, and 
whether it even makes physical sense to scale the mechanical feedback energy 
by the black hole Eddington luminosity. In spite of these uncertainties, 
this test provides a valuable independent  test for the existing models 
of galaxy formation and evolution because of the two 
strong and testable constraints described above. 

Whether the above results are reconcilable with the radio AGN feedback scenario 
in galaxy formation theory is unclear. A na\"ive qualitative prediction based on
these scenarios would be that radio AGN activity should be anticorrelated with 
star formation rate, even at fixed galaxy stellar mass. However, the 
predicted strength of this correlation is not obvious without specifying 
the theoretical scenario more completely. For example, if the time scale 
for the feedback to take effect is long, and the  radio AGN activity 
varies on relatively short time scales, the predicted  correlation between 
star formation and observed radio activity could be rather weak, in agreement 
with observations. 

Therefore, the lack of correlation between radio AGN activity and sSFR could 
reflect the differences in timescales of the two phenomena, or it could indicate
that radio AGN feedback does not play a major role in quenching of massive galaxies. 
Alternatively, it could mean something entirely different. 

To shed light on the origin of the above results, in this study we investigate
the predictions of galaxy formation simulations that incorporate AGN feedback, to
ask if they can be consistent with the observations.
We use the cosmological simulations IllustrisTNG100 (\citealt{illustristng2019}), SIMBA (\citealt{simba}) and EAGLE (\citealt{eagle2015}),
comparing their results to the above observational constraints and evaluating 
their underlying subgrid physics.

Although the resolution of these simulations extends to better than
$1 ~\rm kpc$, they are nowhere near high enough dynamic range to simulate 
the physics of gas accretion onto AGN. 
Consequently, they rely on subgrid models to account for AGN activity, 
among other things. Even though the role of radio AGN in simulations is
broadly the same in all the simulations (to suppress SFR at high $M_{\star}$), 
the details of the AGN subgrid models vary significantly among them. 
Cosmological simulation studies like EAGLE, SIMBA, IllustrisTNG 
etc. have distinct AGN feedback models, and this could give rises to varied 
relationships between radio AGN activity and their host galaxies, which
are testable predictions that might distinguish between these various models.

In this study, we investigate the above mentioned simulations and their
predictions for the relationship between radio AGN activity and host galaxy 
$M_\star$ and sSFR and compare to
the above observational constraints. In the analysis below, we 
will remain cognizant of the systematic uncertainties in the Eddington ratios
and in the connection between radio luminosity and feedback activity. That is,
we do not seek a perfect match between simulations and observations, but 
instead seek to determine whether any reasonable relationship in the simulations
between AGN feedback power and radio luminosities can reproduce our qualitative
results.

In Section \ref{sec:Fagn_constraint} we revisit the observational work from
\cite{sureshblanton2024} and briefly describe an update to the 
$F_{\rm AGN}$ constraint. In Section \ref{sec:cosmo_sims} we discuss the 
cosmological simulations and their black hole feedback models. 
In Section \ref{sec:d&m}, we describe our datasets and analysis methods.
In Section \ref{sec:results}, we discuss the results of the $F_{\rm AGN}$ 
comparison between the simulations and our observations.  

\section{The $F_{\rm AGN}$ constraint}
\label{sec:Fagn_constraint}

\subsection{ERD Fits in $M_\star$ bins for Star forming and Quiescent galaxies}
\label{ssec:ERD_fits}

\cite{sureshblanton2024} show that the radio AGN Eddington ratio distribution 
(ERD) is independent of host sSFR at a given $M_\star$, by showing that 
an sSFR-independent model predicts the observed Eddington ratio distributions
accurately. To improve on this relatively indirect approach and provide a 
better quantification of the uncertainties in this result, here we perform 
a reanalysis of the data.

\begin{figure*}[h]
    \centering
    \includegraphics[width = 7.2in]{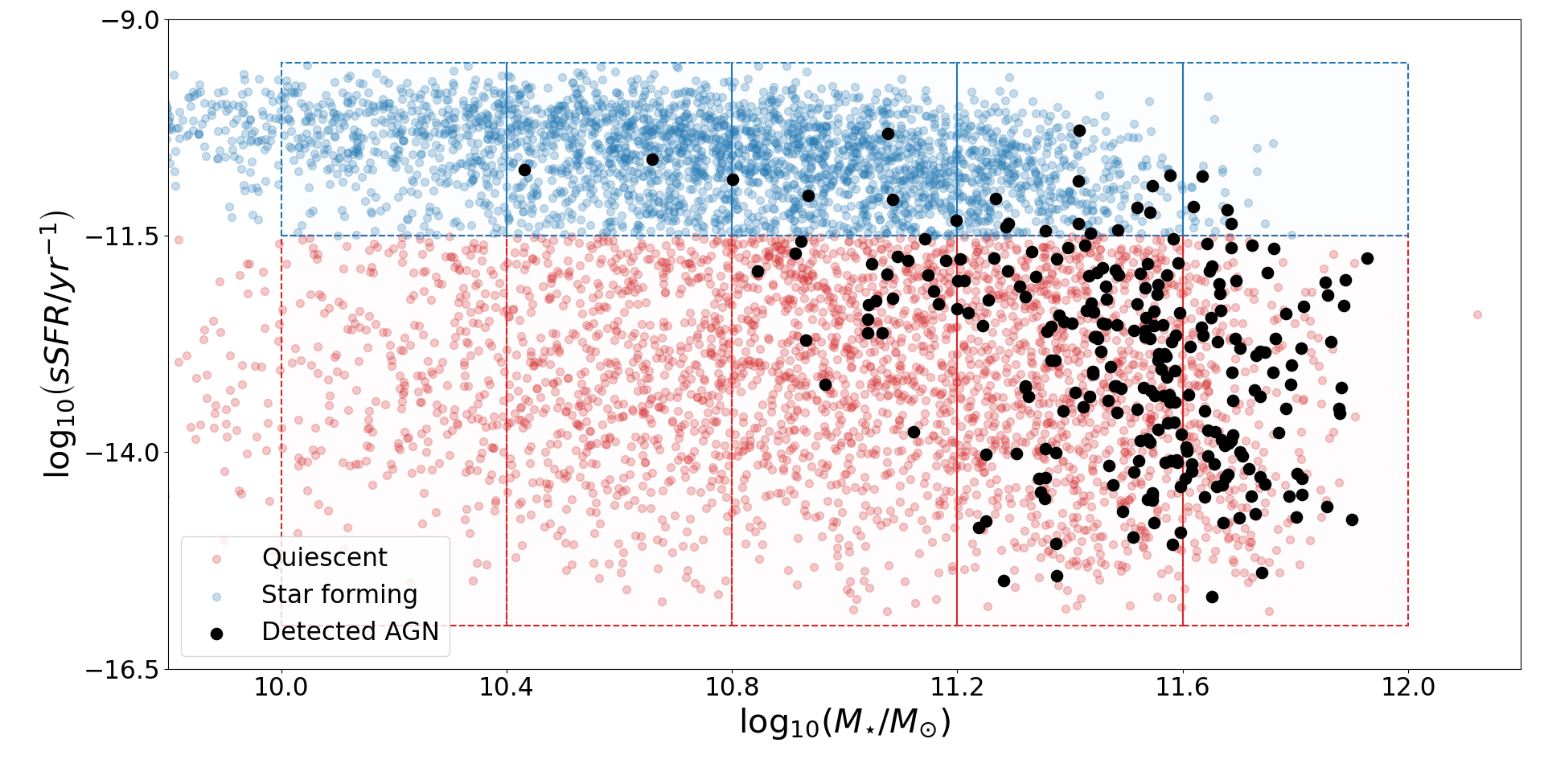}
    \caption{Distribution of galaxies from our MaNGA sample in the $\log_{10}(\mathrm{sSFR}/\mathrm{yr}^{-1})$ versus $\log_{10}(M_{\star}/M_{\odot})$ plane, following the dataset and methodology described in \citet{sureshblanton2024}. Star-forming galaxies are shown in blue, while quiescent galaxies are shown in red, with the dividing line set at $\log_{10}(\mathrm{sSFR}/\mathrm{yr}^{-1}) = -11.5$. Black circular points represent galaxies hosting AGN, identified using the procedure detailed in Section 3.2 of \citet{sureshblanton2024}. Within the stellar mass and sSFR bins outlined by the blue and red dashed rectangles, we perform fits to estimate the radio AGN ERD.}
    \label{fig:ssfr_v_mass}
\end{figure*}

We divide the MaNGA sample into star-forming (SF; 
$ \log_{\rm 10} \left({\rm sSFR}/ {\rm yr}^{-1}\right) > -11.5$) and quiescent galaxies 
(Q; $\log_{\rm 10} \left( {\rm sSFR}/ {\rm yr}^{-1}\right) < -11.5$). We further 
split the sample in bins of $M_\star$, with bin edges at $\log_{\rm 10} 
\left( M_{\star}/ M_{\odot} \right) = [10, 10.4, 10.8, 11.2, 11.6, 12]$. See 
Figure \ref{fig:ssfr_v_mass} for the distribution and binning of the galaxy 
sample in $\log_{\rm 10} \left( {\rm sSFR}/ {\rm yr}^{-1}\right)$ and  
$\log_{\rm 10} \left(M_{\star}/ M_{\odot} \right)$. The big black circles 
are the galaxies that are identified as detected AGN following the method 
described in Section 3.2 of \cite{sureshblanton2024}. The detected AGN have 
measured values for $\lambda$. The remaining galaxies are either radio non-detections
or do not pass our criteria for AGN; these galaxies all have Eddington ratio upper 
limits $\lambda_{\rm lim}$, based on the redshift, the star formation rate of the 
galaxies, and the 2.5 mJy flux limit of NVSS. These sets of $\{ \lambda\}$ and 
$\{ \lambda_{\rm lim}\}$ form the data for the ERD fits. 

In each bin,  following the methods outlined in Section 3.4 of \cite{sureshblanton2024}, 
we fit models for the ERD, using the upper limits to account for the flux and 
star formation related selection effects. We use a Schechter function model, defined
by an exponential cutoff $\lambda_\ast$, a power-law faint-end slope $\alpha$ (positive
values indicate a decline with luminosity), and a minimum Eddington ratio $\lambda_{\rm min}$.
We assume that in a given bin, these parameters  are independent of host galaxy properties 
in that bin. For a detailed  description of the methods, please 
refer to \cite{sureshblanton2024}.

We then evaluate the fits in each bin to constrain the $F_{\rm AGN} 
\left( M_\star\right)$  trends for SF and Q galaxies, using an Eddington
ratio threshold $\lambda > \lambda_{\rm c} = 10^{-3}$. We calculate the 
uncertainties using the posterior distribution of $F_{\rm AGN}$ under the
model, which uses uniform priors in $\alpha$, $\log_{10} \lambda_\ast$ and $\log_{10} \lambda_{\rm min}$. 
However, the lowest mass SF  bin and the two lowest mass Q bins contain no detected 
AGN, and in these cases the prior strongly dominates the $F_{\rm AGN}$ distribution;
because of the nature of the prior volume, leading to artificially strong 
upper limits on $F_{\rm AGN}$.  Therefore, for those three bins we 
calculate model independent upper limits on $F_{\rm AGN}$ with 
$\lambda > \lambda_{\rm c} = 10^{-3}$. The upper limit on \( F_{\rm AGN} \) in each bin is given by
\[ F_{\rm AGN}^{\rm UL} = 1 - \left(0.05\right)^{1/N_{\rm gal}},
\]
where \( N_{\rm gal} \) is the number of galaxies in the bin with \( \lambda_{\rm lim} < 10^{-3} \). This expression corresponds to the 95\% confidence upper limit on the probability of detecting zero AGN, assuming a true AGN fraction of \( F_{\rm AGN} \).

\begin{figure*}[h]
    \centering
    \includegraphics[width = 7.2in]{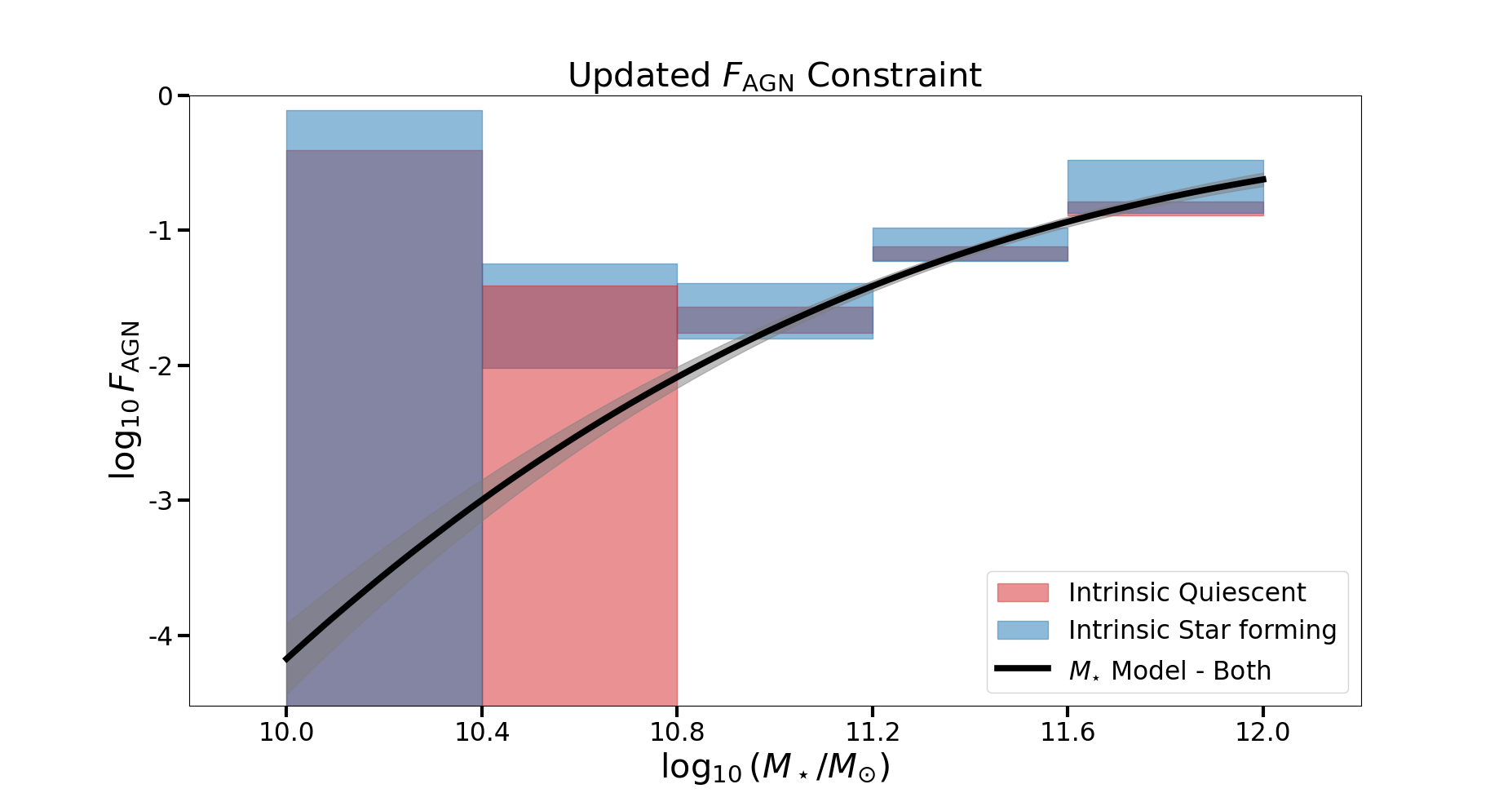}
    \caption{Updated constraints on $F_{\rm AGN}$, derived from the ERD fits described in Section~\ref{ssec:ERD_fits}, are shown within the blue and red rectangular regions. For comparison, the $F_{\rm AGN}$ constraint from \citet{sureshblanton2024} is displayed as a solid black line. The updated results are in excellent agreement with the previous constraint at high stellar masses ($\log_{10}(M_{\star}/M_{\odot}) > 11$). However, at lower stellar masses, we obtain only upper limits on $F_{\rm AGN}$. While these upper limits remain consistent with the earlier results, a larger galaxy sample would be required to reduce uncertainties and improve the statistical precision of the constraints.}
    \label{fig:Fagn_mbins}
\end{figure*}

The updated $F_{\rm AGN}$ constraint is shown in Figure \ref{fig:Fagn_mbins} along with the previous $F_{\rm AGN}$ constraint from \cite{sureshblanton2024}. As seen in the figure, the updated constraint successfully reproduces the observed sSFR independence at high stellar masses $\left(\log_{10}(M_\star / M_\odot) > 11\right)$. At the low-mass end, however, we now obtain only upper limits on \(F_{\rm AGN}\), in contrast to the previous constraint. We believe the earlier, stronger constraint on \(F_{\rm AGN}\) was primarily a consequence of the rigidity of the \(M_\star\)-dependent model of \cite{sureshblanton2024}. While the \(M_\star\)-dependent model was well constrained at high masses due to the abundance of detected AGN, at low masses it was constrained mainly by the model’s rigidity rather than by strong observational data. Although the upper limits remain consistent with the previous constraint, a larger galaxy sample would be necessary to improve the precision; 
in particular, in the two lowest mass bins, we cannot exclude a dependence on sSFR.

We also note that the quiescent bin centered on $\log_{10} M_{\ast} = 10.6$ has only two detected AGN. 
Both of those AGN are high ellipticity galaxies ($\rm e > 0.95$) and lie very close to the 
somewhat arbitrary AGN identification line of $\log_{\rm 10} \left(L_{1.4~ \mathrm{GHz}}/ \mathrm{erg ~ s^{-1}} \right) = 38.6$ (see Figure 2 of \cite{sureshblanton2024}). Based on these reasons, 
we do not entirely trust the classification of the two galaxies as detected AGN. 

For improved clarity about all the above concerns, it would be necessary to perform a 
similar analysis with larger datasets like the Dark Energy Spectroscopic Instrument Bright Galaxy Survey 
(\citealt{DESI, DESIbgs}) or the SDSS Legacy Survey (\citealt{SDSSlegacy}).

We will use this updated constraint to compare the $F_{\rm AGN} \left( M_\star\right)$ predictions from the simulations and evaluate the underlying theories (see Section \ref{sec:results}). We present the updated constraints in Table~\ref{tab:Fagn_bins} to facilitate their use in future studies by readers interested in incorporating these measurements into their own analyses.

\begin{table}[ht]
\centering
\caption{Median values of $\log F_{\mathrm{AGN}}$ and corresponding $1\sigma$ binomial uncertainties are reported in stellar mass bins for both quenched (Q) and star-forming (SF) galaxies. In bins with no AGN detections, we provide the 95\% confidence upper limits on $F_{\mathrm{AGN}}$, assuming a binomial distribution for the underlying probability of AGN occurrence.
}
\begin{tabular}{ccc}
\hline
$\log_{\rm 10}\left(M_\star/M_\odot \right)$ Bin & $\log_{\rm 10} F_{\mathrm{AGN}}$ (Q) & $\log_{\rm 10} F_{\mathrm{AGN}}$ (SF) \\
\hline
$10.0$--$10.4$ & $< -0.41$                      & $< -0.11$ \\
$10.4$--$10.8$ & $< -1.41$                      & $-1.62^{+0.37}_{-0.40}$ \\
$10.8$--$11.2$ & $-1.66^{+0.09}_{-0.10}$        & $-1.58^{+0.19}_{-0.22}$ \\
$11.2$--$11.6$ & $-1.17^{+0.05}_{-0.05}$        & $-1.10^{+0.12}_{-0.13}$ \\
$11.6$--$12.0$ & $-0.84^{+0.05}_{-0.05}$        & $-0.65^{+0.17}_{-0.22}$ \\
\hline
\end{tabular}
\label{tab:Fagn_bins}
\end{table}

\subsection{The Importance of Accounting for Selection Effects}
\label{ssec:selection_effects}

\begin{figure*}[h]
    \centering
    \includegraphics[width = 7.2in]{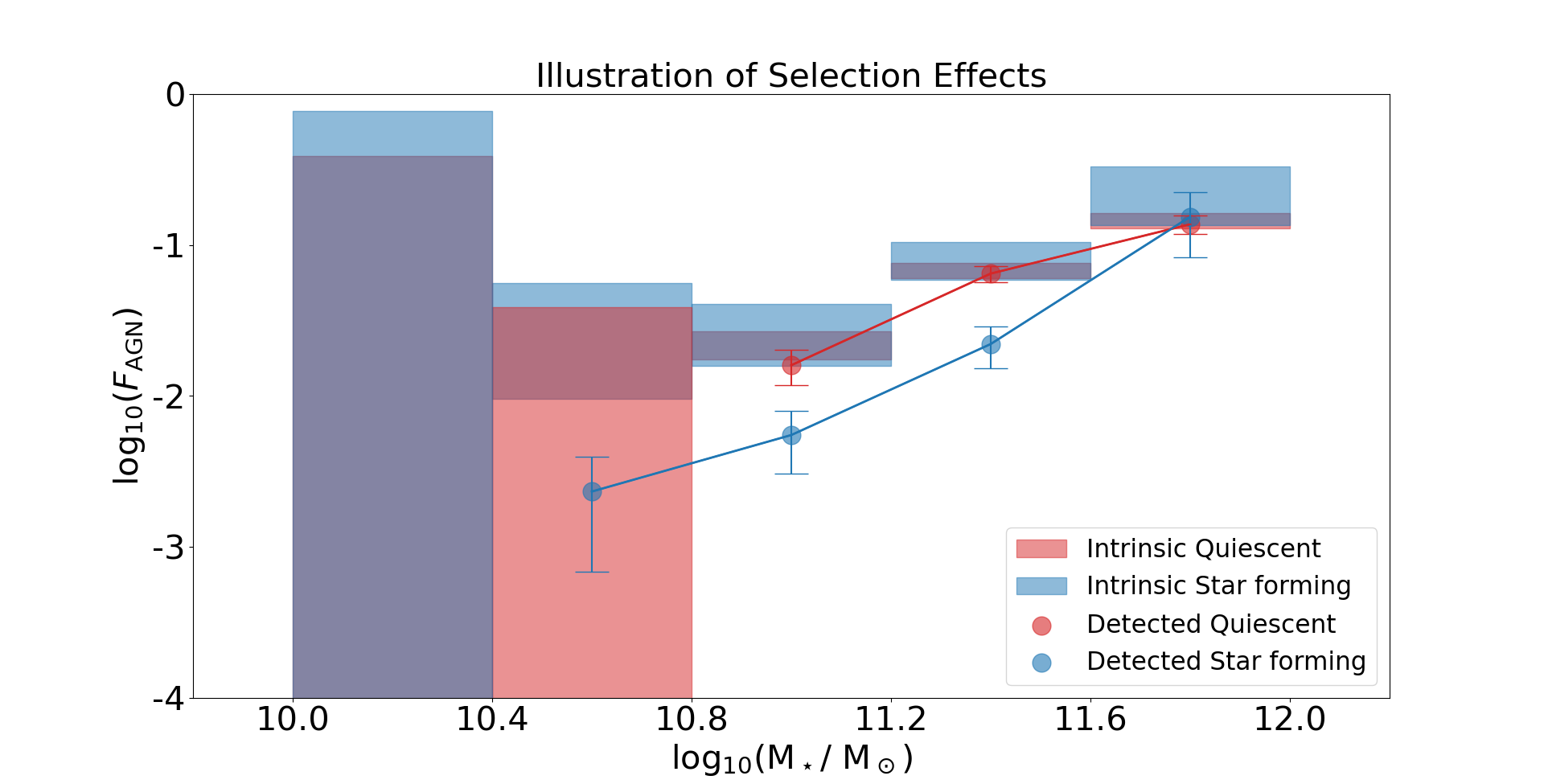}
    \caption{Detected AGN fractions for quiescent and star forming galaxies as seen in our MaNGA sample. The figure also shows the intrinsic AGN fractions for the two populations as evaluated by our modeling of the radio AGN ERD. It highlights the strong difference in the dependence of $F_{\rm AGN}$ on $M_\star$ and $\mathrm{sSFR}$ between the observed and intrinsic populations, clearly demonstrating the importance of accounting for selection effects in AGN studies.}
    \label{fig:selection_effects}
\end{figure*}

The primary advance of \citet{sureshblanton2024}---as well as the fits presented here---beyond 
earlier studies \citep[e.g.,][]{janssen12a, best05a, best12a} lies in the careful 
treatment of relevant selection effects. Specifically, we recognize that 
the AGN--host galaxy relationship observed in the detected AGN population can 
differ significantly from the intrinsic relationship within the full underlying 
population. Consequently, when comparing simulation-predicted AGN–host relations 
to observations, it is crucial that the observational constraints reflect the intrinsic 
population rather than the detected subset. Figure~\ref{fig:selection_effects} 
demonstrates how markedly these two relationships differ in this sample, 
underscoring the importance of properly accounting for selection effects.
The figure compares the observed trends of \( F_{\rm AGN}(M_\star) \) for the detected sample with the intrinsic trends predicted by our model for Eddington ratios \( \lambda > \lambda_{\rm c} = 10^{-3} \).
The two differ significantly, 
illustrating the impact of selection effects. Relying on the detected trend 
alone---as has been done in previous studies (eg: \citealt{best05a,comerford2020,jin2025,chen2013})---could 
lead to the conclusion that radio AGN are preferentially found in quiescent 
galaxies. However, the intrinsic relationship recovered from our model shows 
that this interpretation is clearly inaccurate. 

\section{Cosmological Simulations and their black hole feedback models}
\label{sec:cosmo_sims}

\subsection{Common Definitions}
\label{ssec:common_defs}

Here we describe some definitions and expressions that are very similar across the EAGLE, SIMBA and TNG100 simulations. All of the three simulations incorporate some form of Bondi accretion for black hole growth (\citealt{bondi1952, bondi_hoyle1944, hoyle_littleton_1939}) with a maximum value equal to the Eddington accretion rate $\dot{M}_{\rm Edd}$. These quantities are given by:
\begin{eqnarray}
    \dot{M}_{\rm Bondi} &=& \epsilon_{\rm m} \frac{4 \pi G^2 M_{\rm BH}^2 \rho}{\left( v^2 + c_{\rm s}^2 \right)^{3/2}} \\
    \dot{M}_{\rm Edd} &=& \frac{4 \pi G M_{\rm BH} m_{\rm p}}{\epsilon_{\rm r} \sigma_{\rm T} c}
\end{eqnarray}
where $G$ is Newton's gravitational constant, $M_{\rm BH}$ is the mass of the black hole, $\rho$ is the gas density in the vicinity of the black hole, $c_{\rm s}$ is the local sound speed, and $v$ is the relative speed between the black hole and the gas. Further, $m_{\rm p}$ is the mass of a proton, $\sigma_{\rm T}$ is the Thompson cross section, $c$ is the speed of light and $\epsilon_{\rm r}$ is the radiative efficiency of accretion. Finally, $\epsilon_{\rm m}$ is a normalization constant that encapsulates the efficiency of gas transport from the accretion disk onto the black hole. This quantity is only explicitly included in SIMBA but for consistency we include it here. For the different simulations 
we set $\epsilon_{\rm m}$ and $v$ as follows:
\begin{itemize}
    \item For EAGLE: $\epsilon_{\rm m}$ is fixed to unity.
    \item For SIMBA: The SIMBA team calibrates $\epsilon_{\rm m}$ to have the value 0.1 (see Section \ref{sssec:simba_bh_model}).
    \item For TNG100: $\epsilon_{\rm m}$ is fixed to unity and $v$ is set to 0 (i.e. the relative velocity between the gas and black hole is ignored).
\end{itemize}

The black hole mass accretion rate $\dot{M}_{\rm BH}$ depends on $\dot{M}_{\rm Bondi}$ in different ways for the three simulations, as 
we detail in Sections \ref{sssec:eagle_bh_model}, \ref{sssec:simba_bh_model}, \ref{sssec:tng100_bh_model}. In all three simulations the amount of energy that a black hole outputs depends on $\dot{M}_{\rm BH}$ and is given by:

\begin{equation}\label{eq:edot}
    \dot{E} = K \dot{M}_{\rm BH}c^2
\end{equation}

where,

\begin{itemize}
    \item For EAGLE: $K = \epsilon_{\rm f} \epsilon_{\rm r}$ with $\epsilon_{\rm r} = 0.1$ and $\epsilon_{\rm f} = 0.15$
    \item For SIMBA: $K = \epsilon_{\rm r} = 0.1$ for both quasar and radio mode feedback.
    \item For TNG100: $K = \epsilon_{\rm f} \epsilon_{\rm r}$ for the quasar mode and $K = \epsilon_{\rm kin}$ for the radio mode. $\epsilon_{\rm r} = 0.2$, $\epsilon_{\rm f} = 0.1$ and $\epsilon_{\rm kin}$ is a function of $\rho$ (see Section \ref{sssec:tng100_bh_model}).
\end{itemize}

Here, $\epsilon_{\rm f}$ denotes the fraction of radiated energy that couples with the ISM and for TNG100, $\epsilon_{\rm kin}$ represents the coupling efficiency in the radio mode. Table \ref{tab:parameter_summary} summarizes the values for the above described parameters.

\begin{table*}[t]
\centering
\caption{Summary of analogous black hole feedback parameters across the simulations}
\label{tab:parameter_summary} 
\begin{tabular}{lccccc}
\toprule
Simulation & $\epsilon_{\rm m}$ & $\epsilon_{\rm r}$ & 
$\epsilon_{\rm f}$ & $\epsilon_{\rm kin}$ & $\dot{E}$ \\
\midrule
EAGLE  & $1$ & $0.1$ & $0.15$ & -- & $\epsilon_{\rm f} \epsilon_{\rm r} \dot{M}_{\rm BH} c^2$ \\
\addlinespace
SIMBA  & $0.1$ & $0.1$ & -- & -- & $\epsilon_{\rm r} \dot{M}_{\rm BH} c^2$ \\
\addlinespace
TNG100 & $1$ & $0.2$ & $0.1$ & depends on $\rho$ & 
\makecell[l]{
    Quasar mode: $\epsilon_{\rm f} \epsilon_{\rm r} \dot{M}_{\rm BH} c^2$ \\
    Radio mode: $\epsilon_{\rm kin} \dot{M}_{\rm BH} c^2$
} \\
\bottomrule
\end{tabular}
\end{table*}

\subsection{EAGLE}\label{ssec:eagle}
\subsubsection{EAGLE Cosmological Simulation}\label{sssec:eagle_cosm_sim}
The Evolution and Assembly of GaLaxies and their Environments (EAGLE; \citealt{schaye2015}, \citealt{crain2015}) project is a suite of cosmological hydrodynamic simulations. This suite consists of simulations whose box lengths lie between 25 and 100 comoving Mpc (cMpc), with the largest simulation containing $2 \times 1504^{3}$ particles. The simulations track the evolution of dark matter, gas, stars, black holes between z = 127 and z = 0 and are evolved by using a modified version of the smoothed particle hydrodynamics code GADGET (\citealt{springel_2005}). All of the simulations in the suite assume a flat $\Lambda$ cold dark matter ($\Lambda \rm CDM$) model of the universe, with the parameters $\Omega_{\rm b} = 0.04825$, $\Omega_{\rm m} = 0.307$, $\Omega_{\rm \Lambda} = 0.693$ and $H_{\rm 0} = 67.77$ km~s$^{-1}$~Mpc$^{-1}$. 

For this study, we consider only the Ref-L0100N1504 simulation, which is the primary 
run of the EAGLE suite. It has $2 \times 1504^{3}$ particles and a box length of 100 cMpc, a gas mass resolution $m_{\rm g} = 1.81 \times 10^{6} ~M_{\odot}$ and a dark matter mass resolution $m_{\rm dm} = 9.70 \times 10^{6} ~M_{\odot}$.  This simulation, like the others in the suite, employs subgrid models for black hole growth, gas accretion onto black holes, black hole mergers, AGN feedback, star formation, star formation feedback, radiative cooling, stellar mass loss and metal enrichment. 

The EAGLE team asserts that their feedback subgrid models are simpler and more natural than the feedback subgrid models of other contemporary simulations (\citealt{schaye2015}). Both the stellar and AGN feedback models employ only  a single mode of thermal feedback each to capture the effects of the various processes involved, as opposed to modeling those processes individually. Here, the single mode of stellar feedback aims to effectively capture the effects of stellar winds, radiation pressure on dust grains, and supernovae. Similarly, the single mode of AGN feedback aims to capture the effects of both radiatively efficient and radiatively inefficient AGN activity. The EAGLE team believes that the above processes cannot be distinguished from each other at length resolutions of $10^{2}$--$10^{3} ~\rm pc$, which is typical of the simulations in question. Furthermore, the stellar and AGN feedbacks are injected in the form of thermal energy without turning off radiative cooling and hydrodynamical forces. This in turn generates galactic winds without the wind direction, velocity or mass metal loading factor being specified {\it a priori}. 

\subsubsection{EAGLE's Black Hole Feedback Model}\label{sssec:eagle_bh_model}

Here we summarize the black hole feedback model of the EAGLE Ref-L0100N1504 simulation. This feedback model uses a single mode of feedback where the black holes stochastically inject thermal energy back into their host galaxies. \cite{crain2015} claims that this single mode of feedback mimics the quasar mode or the radio mode of feedback, based on the accretion rate (\citealt{McCarthy10}, \citealt{McCarthy11}). 
The rate of energy injection is given by Equation \ref{eq:edot}.

The black hole mass accretion rate $\dot{M}_{\rm BH}$ is given by: 
\begin{equation}\label{eq:eagle_mdotaccr}
    \dot{M}_{\rm BH} = \min \left(\dot{M}_{\rm Bondi} [(c_{\rm s}/ V_{\phi})^{3}/C_{\rm visc}], \dot{M}_{\rm Bondi} \right)    
\end{equation}
Here, $V_{\phi}$ represents the circular speed of gas around the black hole and $C_{\rm visc}$ is a parameter that quantifies the viscosity of gas. 

Under this model, the black hole has an imaginary reservoir of energy $E_{\rm BH}$ whose value increases by $\epsilon_{\rm f} \epsilon_{\rm r} \dot{M}_{\rm BH} c^{2} \Delta t$ after every time step $\Delta t$. Once $E_{\rm BH}$ has reached a value that is large enough to heat at least one gas element of mass $m_{\rm g}$, the black hole becomes eligible to stochastically choose a neighboring gas particle to raise its temperature by $\Delta T_{\rm AGN}$. The probability that a given neighbor gets heated is:
\begin{equation}
    P = \frac{E_{\rm BH}}{\Delta \epsilon_{\rm AGN} N_{\rm ngb} \langle m_{\rm g} \rangle},
\end{equation}
where $\Delta \epsilon_{\rm AGN}$ is the increase in internal energy per unit mass corresponding to the increase in temperature $\Delta T$, $N_{\rm ngb}$ is the number of gas neighbors of the black hole and $\langle m_{\rm g}\rangle$ is the mean mass of the  neighbors. After injection of the energy, the value of $E_{\rm BH}$ is reduced by the expectation value of the injected energy. 

The EAGLE feedback model has three free parameters, namely, $\epsilon_{f}$, $\Delta T_{\rm AGN}$, and $C_{\rm visc}$. These parameters are tuned in order to calibrate the simulation to observations of the real universe. The observations chosen for calibration are the galaxy stellar mass function (\citealt{li&white2009}, \citealt{baldry12}) and the size-mass relationship of spiral galaxies (\citealt{shen2003}, \citealt{baldry12}), both at z = 0.1. On calibration, the adopted values for the parameters are: $\epsilon_{f} = 0.15$, $\log_{10} \left(\Delta T_{\rm AGN}/ \rm K\right) = 8.5$ and $C_{\rm visc} = 2\pi$. For a more detailed review of the model and its calibration, see \cite{crain2015} and \cite{schaye2015}.

\subsection{SIMBA}\label{ssec:simba}
\subsubsection{SIMBA Cosmological Simulation}\label{sssec:simba_cosm_sim}

SIMBA is a set of cosmological simulations developed with an adapted version of the GIZMO gravity and hydrodynamics solver, operating in the Meshless Finite Mass mode (\citealt{simba} and references within). For this analysis, we use the flagship simulation, which has a box length of 100 $h^{-1}$ cMpc ($h = 0.68$) and uses $2 \times 1024^{3}$ particles, has a gas mass resolution of $m_{\rm g} = 1.2 \times 10^{7}~ M_\odot$ and a dark matter mass resolution of $m_{\rm dm} = 9.6 \times 10^{7}~ M_\odot$. The simulation is run between z = 249 and z = 0, with $\Omega_{\rm m} = 0.3$, $\Omega_{\rm \Lambda} = 0.7$, $\Omega_{\rm b} = 0.048$ and, $H_{0} = 68~ \rm km~s^{\rm -1}~\rm Mpc^{\rm -1}$. 

SIMBA employs subgrid models for star formation, stellar feedback, chemical enrichment, metal loaded winds, radiative cooling, black hole growth, and black hole feedback. The feedback models are partly informed by examination of zoom-in simulations such as the Feedback In Realistic Environments (FIRE; \citealt{FIRE2014, FIRE2018, muratov2015}). 
Whereas all other contemporary simulations use only Bondi accretion to grow their black holes, SIMBA adds an additional mode called ``torque-limited accretion'' (\citealt{hopkins&quataert11}, \citealt{alcazar17a}). The black holes accrete cold gas ($\rm T < 10^{\rm 5}~ \rm K$) via 
torque-limited accretion and hot gas via Bondi accretion ($\rm T > 10^{5} ~\rm K$), 
which SIMBA asserts is a physically appropriate choice. 
The torque-limited accretion model is informed by a combination of analytic 
models and zoom-in simulations of the descent of gas from galactic scales to 
parsec scales, by careful tracking of the angular momentum loss that is necessary. 
The simulations that employ only Bondi accretion for black hole growth ignore 
this key phenomenon and hence unavoidably have to assume black hole self-regulation 
via feedback in order to reproduce the observed black hole--stellar mass 
scaling relations. The torque-limited accretion model mitigates the need for 
such self-regulation, as it reportedly accurately captures the physics of 
the angular momentum redistribution of infalling gas. Of course, the black hole 
feedback in SIMBA is still responsible for driving out gas on galactic scales 
in order to limit black hole growth, but just not through a nonlinear feedback 
loop that Bondi accretion would necessitate (see \citealt{alcazar17a} for more 
information). 

SIMBA's black hole growth parameters are tuned reproduce the observed galaxy stellar mass function at z = 0 (\citealt{bernardi17}) and the observed $M_{\rm BH}-M_{\star}$ relation (\citealt{K&H2013, bentz2018}). In particular, the accretion efficiency parameter $\epsilon_{\rm m}$ is tuned to $10 \%$ to reproduce the observations (see the following subsection). 

SIMBA's black hole feedback model primarily operates in two distinct modes, the high 
accretion rate quasar mode and the low accretion rate radio mode. Both of these modes employ kinetic feedback, with the velocity kicks to gas particles informed by observations of the radio mode and quasar mode in real
AGN. In addition to the two primary modes, 
SIMBA has an additional X-ray mode of feedback that accounts for the energy input 
via X-rays heating and X-ray driven winds. This third mode of black hole feedback uses a combination of kinetic and thermal energy transfer. It is uncertain whether the energy contributions from X-ray mode feedback are directly comparable to our observations. Therefore, for the analysis presented in the text, we exclude the X-ray mode component. However, we performed a separate analysis including the X-ray mode feedback and found that our overall results remain largely unchanged.

\subsubsection{SIMBA's Black Hole Feedback Model}\label{sssec:simba_bh_model}

As stated above, SIMBA employs two main modes of black hole feedback. There is a quasar mode that operates at high Eddington scaled mass accretion rates ($f_{\rm Edd} > 0.02$) and a radio mode that operates at low Eddington scaled mass accretion rates. The radio mode jets begin appearing at $f_{\rm Edd} < 0.2$ but they reach full velocity only at $f_{\rm Edd} < 0.02$.
The Eddington scaled mass accretion rate $f_{\rm Edd} = \dot{M}_{\rm BH}/ \dot{M}_{\rm Edd}$ where $\dot{M}_{\rm BH}$ is the mass accretion rate of the black hole and is given by:

\begin{equation}
    \dot{M}_{\rm BH} = (1 - \epsilon_{\rm r})(\dot{M}_{\rm Torque} + \dot{M}_{\rm Bondi}),
\end{equation}
where $\dot{M}_{\rm Torque}$ is the torque limited accretion rate.

The torque limited accretion rate is given by:
\begin{eqnarray} 
&& \dot{M}_{\rm Torque} =
\epsilon_{\rm T} f_{\rm d} \times \left(\frac{M_{\rm BH}}{10^{8}~ M_\odot} \right)^{1/6}\left(\frac{M_{\rm enc}(R_{0})}{10^{9}~ M_\odot} \right) &\cr
&&\left(\frac{R_{\rm 0}}{100 ~\rm pc} \right)^{-3/2} \left(1 + \frac{f_{\rm 0}}{f_{\rm gas}}\right)^{-1} ~ M_\odot~{\rm yr}^{-1}
\end{eqnarray}

Here, $\epsilon_{\rm T}$ is a normalization factor that encapsulates the efficiency of radial transport of gas from galactic scales onto the black hole and is tuned to have the value 0.1. $R_{0}$ is a radius associated with 
the numerical softening kernel of the black hole, $f_{\rm d}$ is the fraction of baryonic mass within $R_{\rm 0}$ that exists in the form of a disk as measured in the 
simulation, $M_{\rm enc}(R_{0})$ is the total baryonic mass within $R_{\rm 0}$, $M_{\rm BH}$ is the mass of the black hole, $f_{\rm gas}$ is the fraction of the disk mass that is in the form of gas, and:
\begin{equation}
f_{\rm 0} = 0.31f_{\rm d}^2 \left( \frac{M_{\rm d} \left(R_{\rm 0} \right)} {10^9 M_\odot}\right)^{-1/3}
\end{equation}
We only mention this equation here so it is clear which parameters are necessary for the 
simulation to set; for a full description please refer to \cite{simba} and the references 
within.

SIMBA imposes upper limits on $\dot{M}_{\rm Torque}$ and $\dot{M}_{\rm Bondi}$ based on the black hole's $\dot{M}_{\rm Edd}$. While $M_{\rm Bondi}$ has a strict upper limit equal to $\dot{M}_{\rm Edd}$, $\dot{M}_{\rm Torque}$ has an upper limit of $3\dot{M}_{\rm Edd}$, to allow for super Eddington accretion (\citealt{martinez&aldama2018}, \citealt{jiang2014}). 

The power $\dot{E}$ that a given black hole outputs depends on $\dot{M}_{\rm BH}$ and is given by Equation \ref{eq:edot}. SIMBA assumes $\dot{E}_{\rm kin}$ out of $\dot{E}$ to be in kinetic form and SIMBA further assumes that it is mostly $\dot{E}_{\rm kin}$ that couples with the galaxy. The kinetic power depends on the outflow velocity of gas $v$ and the mass loading factor $\dot{M}_{\rm out}$ as $\dot{E}_{\rm kin} = \frac{1}{2} \dot{M}_{\rm out} v^{2}$. The outflow velocities of the high $f_{\rm Edd}$ mode winds and low $f_{\rm Edd}$ mode jets depend on $M_{\rm BH}$ and $\dot{M}_{\rm BH}$, and are informed by observed optical spectra of AGN (\citealt{perna17a, fabian2012}). Once the outflow velocity is determined, $\dot{M}_{\rm out}$ is set by using observed scaling relations between momentum output $\dot{P}_{\rm out}=\dot{M}_{\rm out} v$ and the total power output $\dot{E}$ (\citealt{fiore17}). In particular, $\dot{P}_{\rm out}$ depends on $\dot{E}$ as:
\begin{equation}\label{eq:momentum_dot}
    \dot{P}_{\rm out} = \frac{20 \dot{E}}{c}
\end{equation}
The ratio $\dot{P}_{\rm out} / \dot{E}$ is kept constant for both the modes of feedback. Hence the fraction of the total feedback power that couples with the galaxy i.e. the kinetic power $\dot{E}_{\rm kin}$ is set for a black hole given its $M_{\rm BH}$ and $\dot{M}_{\rm BH}$.

At high accretion rates ($f_{\rm Edd} > 0.02$) the outflows are intended to mimic the radiative AGN winds that are seen in the quasar mode feedback. The outflow velocities are given by:
\begin{equation}
v_{\rm w} = 500 + \frac{500}{3}\times( \log \left(M_{\rm BH}/ M_\odot \right) - 6)~ \rm km \rm s^{-1}
\end{equation}

At low accretion rates ($f_{\rm Edd} < 0.2$), the feedback begins to transition into the radio mode and eventually peaks at $f_{\rm Edd} < 0.02$. These outflows are intended to mimic the highly relativistic AGN jets and hence contain an additional jet velocity component. To prevent low mass black holes with a temporary low accretion rate from launching jets, SIMBA employs a black hole mass cut of $M_{\rm BH} > 10^{\rm 7.5} M_{\rm \odot}$ as motivated by observations (\citealt{barisic17}). The jet outflow velocity $v_{\rm jet}$ is given by:

\[
v_{\rm jet} = 
\begin{cases} 
0, & \text{if } M_{\rm BH} < 10^{7.5} M_\odot, \\[8pt]
7000 \cdot \log \left(\frac{0.2}{f_{\rm Edd}} \right) \, \mathrm{km \, s^{-1}}, & \text{if } M_{\rm BH} \geq 10^{7.5} M_\odot.
\end{cases}
\]

The total outflow velocity for a black hole $v$ is the sum of $v_{\rm w}$ and $v_{\rm jet}$. The kinetic power associated with the outflows is given by:

\begin{equation}\label{eq:simba_pkin}
    \dot{E}_{\rm Kin} = \frac{1}{2} \dot{M}_{\rm out} v^{2} = 10\epsilon_{\rm r} c \dot{M}_{\rm BH} v
\end{equation}

Where Equation \ref{eq:momentum_dot} has been used to substitute for $\dot{M}_{\rm out}$.

\subsection{IllustrisTNG100-1}
\subsubsection{IllustrisTNG100-1 Cosmological Simulation}\label{sssec:tng100_cosm_sim}

IllustrisTNG (\citealt{pillepich18}, \citealt{rweinberger2018}) is a suite of cosmological simulations and is the successor of the original Illustris simulation (\citealt{vogelsberger14}). The IllustrisTNG simulations are built using AREPO (\citealt{springel2010}). AREPO is a moving mesh code used to solve the equations of self gravity and magneto-hydrodynamics. The simulations are evolved between z = 127 and z = 0, assuming a $\Lambda$CDM model of the Universe, with parameters taken from the Planck mission (\citealt{planck_collab2014}). The adopted values for the cosmological parameters are: $\Omega_m = 0.3089$, $\Omega_b = 0.0486$, $\Omega_{\rm \Lambda} = 0.6911$ and the dimensionless Hubble constant $h = 0.6774$. 

For this study we make use of the IllustrisTNG100-1 simulation (just TNG100 henceforth), which has $1820^{3}$ gas and dark matter particles each and a box length $L_{\rm box} = 75 \rm ~h^{\rm -1} ~Mpc$. The gas mass resolution $m_{\rm gas} = 1.4 \times 10^{6} M_{\odot}$ and a dark matter mass resolution $m_{\rm dm} = 7.5 \times 10^{6}$. TNG100 employs subgrid prescriptions for star formation, stellar evolution, chemical enrichment primordial and metal line cooling of gas, stellar feedback and black hole formation, growth and feedback. The parameters of the subgrid models have been tuned to match observations in the local Universe. In particular, they are tuned to match the global star formation rate density as a function of cosmic time (\citealt{behroozi18}, \citealt{oesch2015}), the galaxy stellar mass function at z = 0 (\citealt{baldry2008}, \citealt{bernardi2013}) and the stellar mass to halo mass relation at z = 0 (\citealt{behroozi18}, \citealt{moster2013}).

\subsubsection{TNG100's Black Hole Feedback Model}\label{sssec:tng100_bh_model}

Here we summarize TNG100's black hole feedback model (\citealt{rweinberger2018}). This model operates in two distinct modes depending on the Eddington scaled mass accretion rates. At high accretion rates, the feedback mechanism is thermal, and is akin to the observed quasar feedback mode of AGN. At low accretion rates, the feedback mechanism is kinetic and is akin to the radio-mode feedback of AGN. However, physically, the kinetic mode is by design different than the observed radio mode. While observations clearly show that the radio-mode feedback can involve bipolar jets that constitute bulk of the feedback energy, the TNG100 feedback model chooses to have a feedback model that employs kicks in a single direction with the direction changing randomly across injections. Over time, this feedback mechanism tends to be isotropic since the directions are picked at random. This is more like the theorized kinetic winds that accompany jets in the radio-mode feedback. Theoretical work has shown that these winds could play a dominant role in the feedback cycle, more important than the jets themselves. One of the central motivations of TNG100's black hole feedback model is to test the above claim.

In both the modes of AGN activity, the black holes accrete gas via Bondi accretion. The Eddington scaled mass accretion rate in both the modes is given by:
\begin{equation}
\dot{M} = \dot{M}_{\rm Bondi}/ \dot{M}_{\rm Edd} 
\end{equation}

When a given black hole has $\dot{M} \geq \chi$ the black hole operates in the quasar mode and when $\dot{M} < \chi$ it operates in the radio mode. Here, $\chi$ is a mass-dependent transition threshold, in contrast to most other simulations where it is assumed to be a constant. Motivated by observations of X-ray binaries (\citealt{dunn2010}), $\chi$ is allowed to take on a value between 0.001 and 0.1. Since radio mode feedback is more likely to be observed in the very massive black holes, the threshold $\chi$ is allowed to depend on the black hole mass in the following way:

\begin{equation}\label{eq:default_transition_chi}
\chi = \rm min\left(\chi_{\rm0} \left(\frac{M_{\rm BH}}{10^{8} M_{\odot}}\right)^{\beta}, 0.1 \right)
\end{equation}

Here $\beta$ and $\chi_{\rm 0}$ are tunable parameters set to 2 and 0.002 respectively (\citealt{rweinberger2018}). For $\beta > 0$, this sort of dependence of $\chi$ on $M_{\rm BH}$ makes it very hard for the most massive black holes to transition to the quasar mode but still possible if the accretion rate is very high. Furthermore, it makes it very hard for the low mass black holes to operate in the radio mode. The simulators of TNG100 expect this sort of a model to facilitate the rapid quenching of the most massive galaxies via kinetic feedback while leaving the low mass galaxies unaffected by it.

 In the high-accretion or quasar mode, the feedback is in the form of pure thermal energy. In the low-accretion or radio mode, the feedback is in the form of kinetic energy. The energy output of the black holes depends on  $\dot{M}_{\rm BH}$ and the rate of energy injection for the two modes is given by Equation \ref{eq:edot}.

The value of the radio mode coupling efficiency $\epsilon_{\rm kin}$ depends on the density of gas around the black hole and is given by:
\begin{equation}
\epsilon_{\rm kin} = \rm min\left(\frac{\rho}{f_{\rm thresh} \rho_{\rm SF~thresh}}, 0.2\right)
\end{equation}
where $f_{\rm thresh} = 0.05$ and $\rho_{\rm SF~thresh}$ is the threshold gas density for star formation and has a value that is equivalent to $n_{\rm H} = 0.13$ $\rm cm^{-3}$. 

\section{Data and Methodology}\label{sec:d&m}

\subsection{Data and Methodology Overview}\label{ssec:d&m_overview}

In this section we describe the datasets and the methods we use to evaluate and compare the black hole feedback models of the cosmological simulations against the observational constraint. The observational constraint is valid only in the mass regime $10 \le \log_{10}\left( M_{\star}/ M_\odot \right) \le 12$ and we impose this cut for the catalogs of all the simulations. 

We quantify the activity of the black holes in terms of their Eddington ratio. Here, by Eddington ratio, we mean the ratio of a black hole's radio or kinetic mode feedback power ($P_{\rm kin}$) to its Eddington luminosity ($L_{\rm Edd}$). A black hole's $L_{\rm Edd}$ and $\lambda$ depend on its mass ($M_{\rm BH}$) and they are given by: 

\begin{equation}
     L_{\rm Edd} = 1.26\times 10^{38}\times  M_{\rm BH}/ \rm {M_\odot} {\rm ~ergs~s}^{-1} 
\end{equation}

\begin{equation}\label{eq:ER}
    \lambda = P_{\rm Kin}/ L_{\rm Edd}
\end{equation}
    
Since the different simulations in question have vastly different black hole feedback models, the specifics of the definition of $P_{\rm Kin}$ and hence $\lambda$ can vary significantly among them. Due to this, the absolute scale of $P_{\rm Kin}$ and hence of $\lambda$ can vary significantly across individual simulations. Furthermore, our observational constraint
on $P_{\rm Kin}$ is based on radio luminosity, which itself has an uncertain relationship with
any actual feedback power the AGN supplies in the real universe. Keeping in mind this uncertainty in the absolute scales of $\lambda$, we perform the following test to evaluate the black hole feedback models of the simulations in question. 

We compare the predicted $F_{\rm AGN} (M_\star)$ trend by each simulation to the observational constraint in Section \ref{sec:Fagn_constraint}. Here, $F_{\rm AGN}$ is the fraction of galaxies at a given mass that are radio AGN with $\lambda \geq \lambda_{\rm c}$; where the value of $\lambda_{\rm c}$ is fixed at $10^{-3}$ for the observational constraint in Figure \ref{fig:Fagn_mbins}. 

To account for potential inconsistencies in the absolute scale of $\lambda$ between observations and simulations, we perform this comparison with three different choices of $\lambda_{\rm c}$ for each simulation by introducing an additive parameter $\Delta \log_{\rm 10} \left( \lambda_{\rm c} \right)$. That is, for each simulation, for comparison to observations (with a fixed $\lambda_{\rm c}$), we use the theoretical prediction for $F_{\rm AGN} (M_\star)$ 
at $\log_{\rm 10} \left(\lambda_{\rm c}^{\prime}\right) = \log_{\rm 10} \left(\lambda_{\rm c}\right) + \Delta \log_{\rm 10} \left( \lambda_{\rm c} \right)$. We will allow $\Delta \log_{\rm 10} \left( \lambda_{\rm c} \right)$ to take the values -1, 0, and 1.

\subsection{Data and Methodology: EAGLE}\label{ssec:d&m_eagle}

\subsubsection{Data Samples: EAGLE}\label{sssec:data_eagle}

EAGLE has galaxy and halo catalogs available for their simulations suite at 29 snapshots 
between $z = 20$ to $z = 0$ (\citealt{mcalpine2015}). As previously mentioned, we 
consider only the Ref-L0100N1504 simulation of the EAGLE suite and use only the catalogs 
pertaining to snapshots 27 and 28 ($z = 0.1$ and $z = 0$). These two snapshots have 
a total of 7,313 galaxies in the mass range 
$10 \leq \log \left(M_{\star}/ M_\odot \right) \leq 12$. The two snapshots 
contain largely  the same galaxies, just at two different times. Nevertheless,
because the fraction of time each galaxy spends as an AGN is relatively low 
and the variability time scale is less than the billion  years between $z=0.1$ 
and $z=0$, using both allows improves the statistical power in the theoretical
prediction.

Among other quantities, the catalog has estimates available for stellar mass ($M_\star$), black hole mass ($M_{\rm BH}$), black hole mass accretion rate ($\dot{M}_{\rm BH}$), star formation rate (SFR), making the specific star formation rate (sSFR) easy to be computed. The values for $M_{\rm BH}$ in the catalog correspond to the sum of the masses of all the black hole particles in a given galaxy. For $M_{\rm BH} > 10^{6} M_{\odot}$, this approximates closely to the mass of the central supermassive black hole (\citealt{mcalpine2015}), which is the relevant quantitiy for our study. We choose to work with this quantity as our study is concerned with black holes with $M_{\rm BH} \gtrsim 10^{8} ~M_\odot$. We make a similar approximation for $\dot{M}_{\rm BH}$ as well.

\subsubsection{EAGLE: Eddington Ratio Estimates}\label{sssec:ER_eagle}

As previously discussed, EAGLE employs only a single thermal mode of operation for its black hole feedback model. The EAGLE team concluded that as implemented this single mode performed 
both the function of maintenance mode feedback at low mass accretion rates and the 
function of quasar mode feedback at high mass accretion rates, relative to Eddington 
(\citealt{McCarthy10, McCarthy11, crain2015}). 

For each central black hole in the 7,313 galaxies we ascribe an Eddington ratio $\lambda = P_{\rm Kin}/ L_{\rm Edd}$, with $P_{\rm Kin} = \dot{E} = \epsilon_{\rm f} \epsilon_{\rm r} \dot{M}_{\rm BH} c^{2}$ from Equation \ref{eq:edot}. Out of these, we only select the black holes with $ \lambda < 0.02$ as being radio AGN. For the above described EAGLE data, and with $\lambda$ as defined here, we calculate EAGLE's prediction for $F_{\rm AGN}(M_\star)$ with $\Delta \log_{\rm 10} \left( \lambda_{\rm c} \right) = -1,0 ~\text{and}~1$.

In the EAGLE simulation, the condition $\lambda < 0.02$ merely represents a tendency for AGN to exhibit radio-mode-like behavior, rather than explicitly enforcing a distinct radio mode. To assess the implications of not imposing this threshold, we also conducted our analysis without applying this constraint. We find that our results remain largely unaffected by its exclusion.

\subsection{Data and Methodology: SIMBA}\label{ssec:d&m_simba}

\subsubsection{Data Samples: SIMBA}\label{sssec:data_simba}

SIMBA has publicly available galaxy catalogs for 151 snapshots 
between z = 20 and z = 0 (\citealt{simba}). For this analysis, we use the final nine 
snapshots (at $0 \leq z \leq 0.14$) of the flagship run, again using
multiple snapshots to improve statistics. There exist 91,422 galaxies with $ 10 \leq \log \left(M_\star/M_\odot \right) \leq 12$ in the above redshift range. Among other quantities the catalogs provide estimates for $M_\star$, SFR, $M_{\rm BH}$, $\dot{M}_{\rm BH}$ and $f_{\rm Edd}$ allowing us to calculate sSFR in a straightforward manner.

\subsubsection{SIMBA: Eddington Ratio Estimates}\label{sssec:ER_simba}

As previously discussed, SIMBA employs two major modes of feedback -- quasar mode and radio mode. For the central black holes in the sample that are operating in radio mode feedback, we ascribe an Eddington ratio $\lambda = P_{\rm Kin}/ L_{\rm Edd}$. Here, $P_{\rm Kin}  = 10 \epsilon_{\rm r} c  \dot{M}_{\rm BH} v$ from Equation \ref{eq:simba_pkin}. For the above described SIMBA data, and with $\lambda$ as defined here, we calculate SIMBA's prediction for $F_{\rm AGN}(M_\star)$ with $\Delta \log_{\rm 10} \left( \lambda_{\rm c} \right) = -1,0, ~\text{and}~1$.

\subsection{Data and Methodology: TNG100}\label{ssec:d&m_tng100}

\subsubsection{Data Samples: TNG100}\label{sssec:data_tng100}

TNG100 has publicly available galaxy and subhalo catalogs at a hundred snapshots between $z = 20.05$ and $z = 0$. For this study we use the catalogs at $ 0\leq z \leq 0.15$, i.e. snapshot numbers 87 through 99. The galaxy catalog provides us with estimates for the star formation rates (SFR) and stellar masses ($M_{\star}$) of galaxies. There exist 70,820 galaxies that have $10 \leq \log \left( M_{\star}/ M_\odot \right) \leq 12$. Along with the galaxy catalogs, we also make use of the particle level snapshot data for information on the galaxy's super massive black hole. These include the black hole's mass $M_{\rm BH}$, mass accretion rate $\dot{M}_{\rm BH}$ and the gas density $\rho$ in its immediate vicinity.

\subsubsection{TNG100: Eddington Ratio Estimates}\label{sssec:ER_tng100}
As previously mentioned, the black hole feedback model of TNG100 operates in two distinct modes. For the black holes operating in the low Eddington mode or the kinetic mode of feedback we ascribe an Eddington ratio $\lambda = P_{\rm Kin}/ L_{\rm Edd}$. Here $P_{\rm Kin} =  \dot{E} = \epsilon_{\rm kin} \dot{M}_{\rm BH}c^2$  for the radio mode, as given by Equation \ref{eq:edot}. For the above described TNG100 data, and with $\lambda$ as defined here, we calculate TNG100's prediction for $F_{\rm AGN}(M_\star)$ with $\Delta \log_{\rm 10} \left( \lambda_{\rm c} \right) = -1,0, ~\text{and}~1$.


\section{Results}\label{sec:results}

Here we present the results of our comparison between the predicted and observed 
dependencies of $F_{\rm AGN}$ on host galaxy $M_\star$ and sSFR. 
With the potential lack of consistency in the scale of Eddington ratios across the 
observations and simulations in mind, we primarily pay attention to the general trend of 
$F_{\rm AGN}$ with respect to $M_\star$ and lack thereof with respect to sSFR, 
rather than expect agreement in the absolute values. As noted in Sections \ref{intro} and \ref{sec:Fagn_constraint}, we find that a 
galaxy's likelihood of hosting a radio AGN is primarily determined by its 
stellar mass, with its specific star formation rate (sSFR) having minimal or 
no impact.

\subsection{Metrics for comparison} 
\label{sec:metrics}

For each of the simulations in question we generate its prediction for $F_{\rm AGN}(M_{\star})$ separately for star-forming galaxies and quiescent galaxies at $\Delta \log_{\rm 10} \left( \lambda_{\rm c} \right) = -1,0, ~\text{and}~1$. Each of those predictions is then compared with the observational constraint as a means of evaluation of the underlying black hole feedback models.  We evaluate the predictions for the following parameters:

\begin{itemize}
    \item To test for strong $M_\star$ dependence: We approximate the predicted and observed $F_{\rm AGN}(M_{\star})$ curves as straight lines. We then compare the slopes of the models against the observations. The value of the slope for the observed $F_{\rm AGN}(M_{\star})$ curve is $\sim 1.03$ for quiescent galaxies and $\sim 0.85$ for star forming galaxies.
    \item To test for sSFR independence: We use the RMS deviation in the predicted $F_{\rm AGN}(M_\star)$ values between star-forming galaxies and quiescent galaxies. This quantity is $~ 0.23$ dex in our observations. 
\end{itemize}

These parameters are qualitative tests of the model predictions and are not sensitive to quantitative deviations relative to the observations. As we show in the following subsections, none of the simulations in question, for any choice of $\Delta \log_{\rm 10} \left( \lambda_{\rm c} \right)$, predict $F_{\rm AGN}(M_{\star})$ trends that perform satisfactorily on either of the two parameters. We note here that to derive these parameters, we have made use of only those bins in which we have $1\sigma$ confidence intervals. We have ignored the bins in which all we have is an upper limit. 

\subsection{$F_{\rm AGN}$ comparison for EAGLE, SIMBA and TNG100}\label{ssec:results_fagn}

\begin{figure*}[h]
    \centering
    \includegraphics[width = 7.2in]{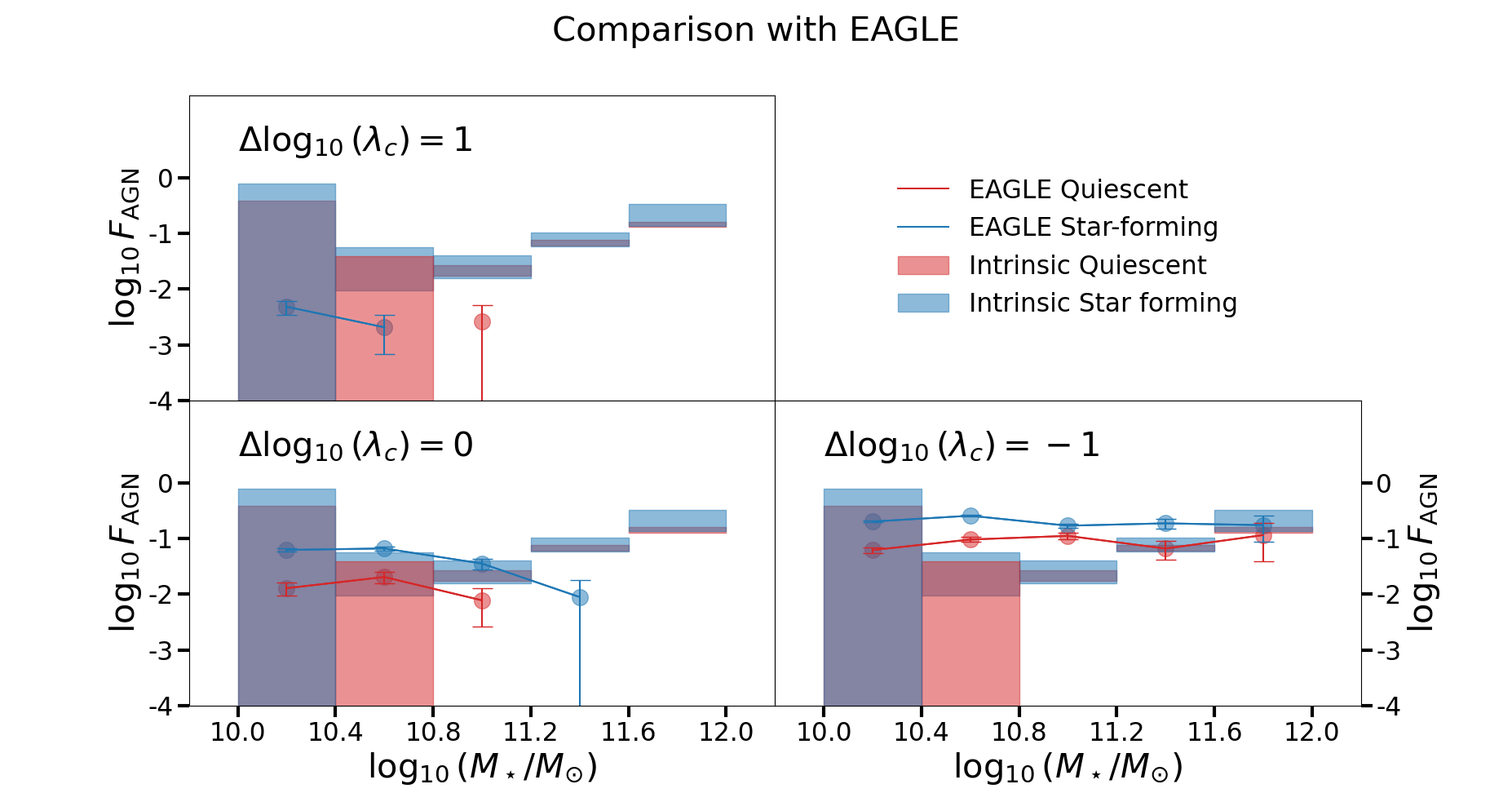}
    \caption{Comparison of the EAGLE simulation's predictions for \( F_{\rm AGN}(M_\star) \), including \( 1\sigma \) binomial uncertainties for both star-forming and quiescent galaxy populations, against the observational constraints shown in Figure~\ref{fig:Fagn_mbins}. Red points represent quiescent galaxies in EAGLE and blue points represent star-forming galaxies in EAGLE, separated at $\log_{10}(\mathrm{sSFR}/\mathrm{yr}^{-1}) = -11$. To account for potential differences in the normalization of $\lambda$ between observations and simulations, we present three subplots corresponding to different values of $\Delta \log_{10}(\lambda_{\rm c})$. The comparison reveals that EAGLE fails to reproduce the observed strong dependence of $F_{\rm AGN}$ on $M_\star$, as well as its relative independence from $\mathrm{sSFR}$.}
    \label{fig:Fagn_eagle}
\end{figure*}

\begin{figure*}[h]
    \centering
    \includegraphics[width = 7.2in]{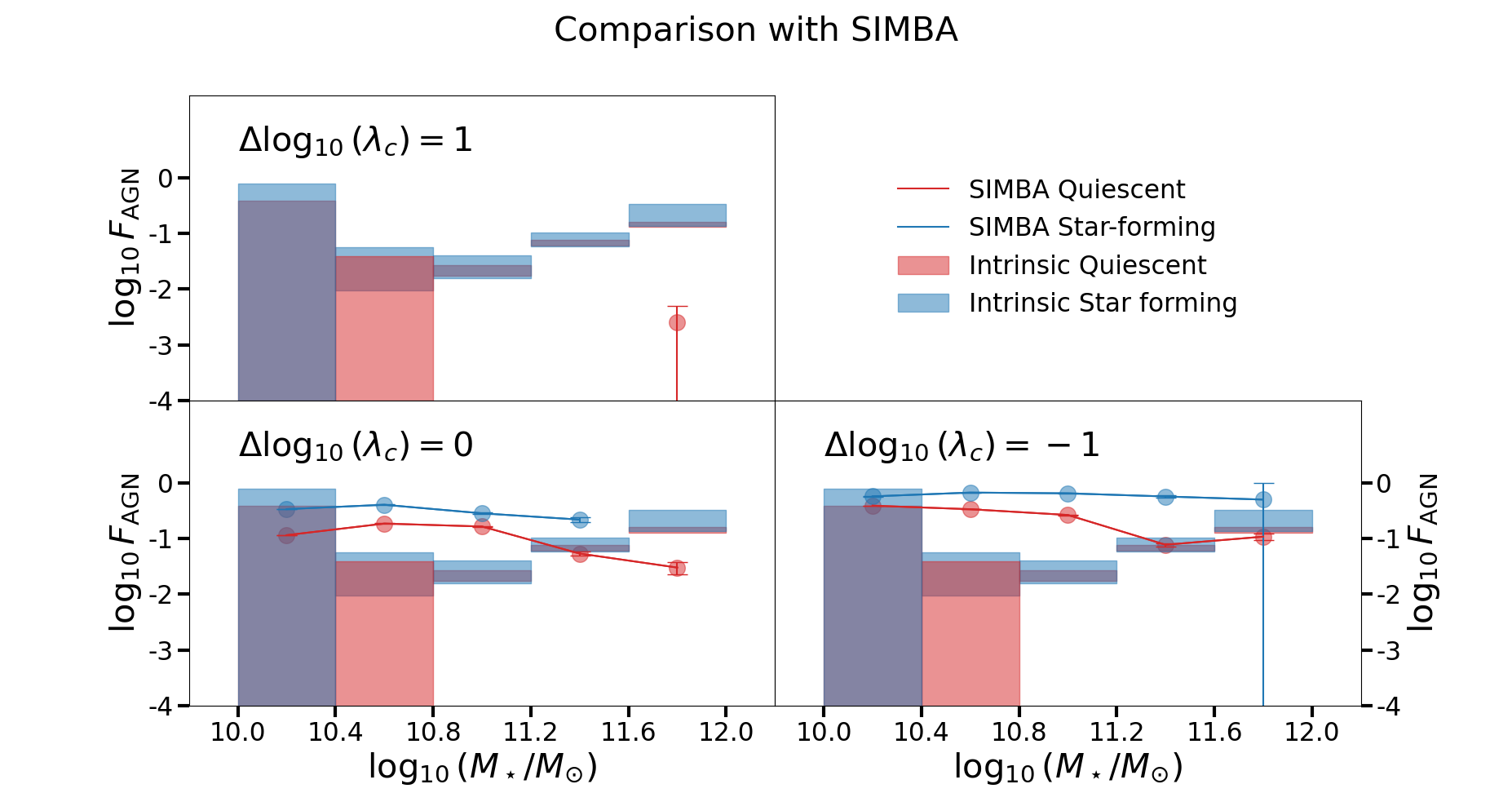}
    \caption{Similar to Figure \ref{fig:Fagn_eagle}, for SIMBA. The dividing line between quiescent and star-forming galaxies in SIMBA is set at \( \log_{10} \left( \mathrm{sSFR}/\mathrm{yr}^{-1} \right) = -11.5 \), consistent with SIMBA's internal classification.}
    \label{fig:Fagn_simba}
\end{figure*}

\begin{figure*}[h]
    \centering
    \includegraphics[width = 7.2in]{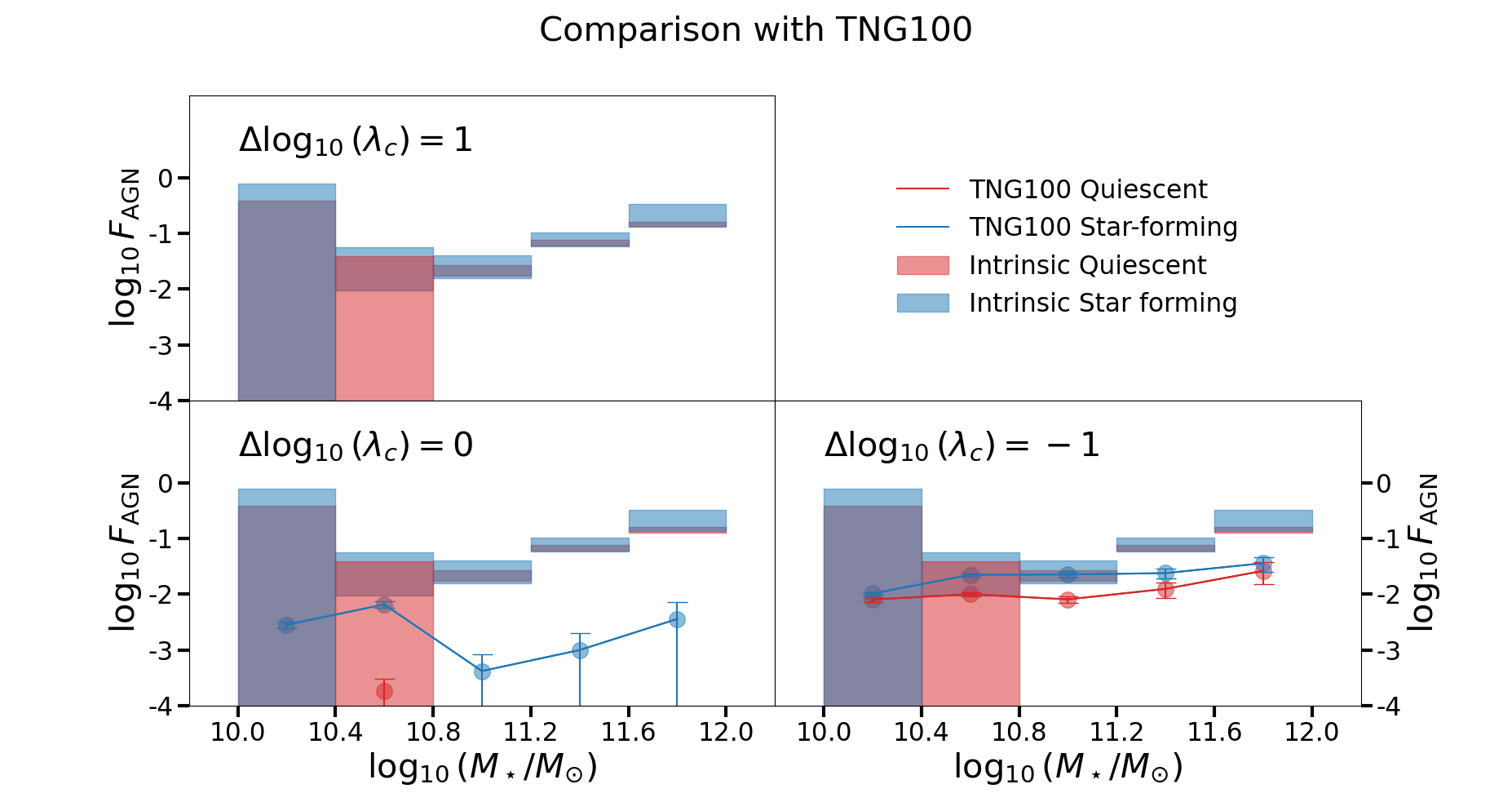}
    \caption{Similar to Figure \ref{fig:Fagn_eagle}, for TNG100. The dividing line between quiescent and star-forming galaxies in TNG100 is set at \( \log_{10} \left( \mathrm{sSFR}/\mathrm{yr}^{-1} \right) = -11.5 \), consistent with TNG100's internal classification.}
    \label{fig:Fagn_illustris}
\end{figure*}

Here we describe the results of the comparison for all the simulations under consideration. The three subplots of Figure \ref{fig:Fagn_eagle} show the results for the EAGLE simulation with each subplot corresponding to a different value of $\Delta \log_{\rm 10}\lambda_{\rm c}$. In each subplot, the blue and red shaded regions denote the observational constraints on $F_{\rm AGN}$ as a function of $M_\star$ for star forming and quiescent galaxies, respectively. EAGLE's predictions for $F_{\rm AGN}$ are shown as blue and red points for star forming and quiescent galaxies, respectively. Galaxies with $\log_{10} \left( {\rm sSFR}/ M_\odot \right) > -11$ are considered to be star-forming and those with $\log_{10} \left( {\rm sSFR} / M_\odot \right) < -11$ are considered to be quiescent, in line with EAGLE's definitions of them (\citealt{crain2015}). The error bars on the red and blue points denote the $1\sigma$ uncertainties around the expectation values, assuming a binomial distribution. The top left plot shows the comparison at $\Delta \log_{\rm 10} \left( \lambda_{\rm c} \right) = 1$ and the two plots on the bottom show the comparison at $\Delta \log_{\rm 10} \left( \lambda_{\rm c} \right) = 0 \rm ~and~ \Delta \log_{\rm 10} \left( \lambda_{\rm c} \right) = -1$. Figures \ref{fig:Fagn_simba} and \ref{fig:Fagn_illustris} show analogous plots for SIMBA and TNG100 respectively.  

The plots clearly show that the simulations fail to perform well against either of the two parameters of the test described in Section \ref{sec:metrics}: 

\begin{itemize}

    \item $M_\star$ dependence: None of the models in question, for any $\Delta \log_{\rm 10} \left( \lambda_{\rm c} \right)$, are able to reproduce the strong $M_\star$ dependence seen in the observations. The slopes of the predicted curves almost always carry the incorrect sign (negative) in relation to the observations (positive). The only exception is TNG100 at $\Delta \log_{\rm 10} \left( \lambda_{\rm c} \right) = -1$, which has a slope 0.64 for the quiescent population and 0.16 for the star forming population. 

    \item sSFR independence: None of the models are able to reproduce the observed sSFR independence down to a precision of 0.23 dex. EAGLE and TNG100 seem to perform reasonably well at $\Delta \log_{\rm 10} \left( \lambda_{\rm c} \right) = -1$ with an RMS deviation of 0.30 (bottom right panel of Figure \ref{fig:Fagn_eagle}) and 0.32 (bottom right panel in Figure \ref{fig:Fagn_illustris}), respectively. These aren't quite the required precision but they are close to it.
\end{itemize}

To summarize, for no choice of $\Delta \log_{\rm 10} \left( \lambda_{\rm c} \right)$ are the models able to reproduce the observed $M_{\star}$ dependence and sSFR independence of $F_{\rm AGN}$. The models perform poorly against both the parameters that we evaluate them on. The predictions show significant qualitative --- and therefore quantitative --- discrepancies compared to the observations. 

\begin{figure*}[h]
    \centering
    \includegraphics[width = 7.2in]{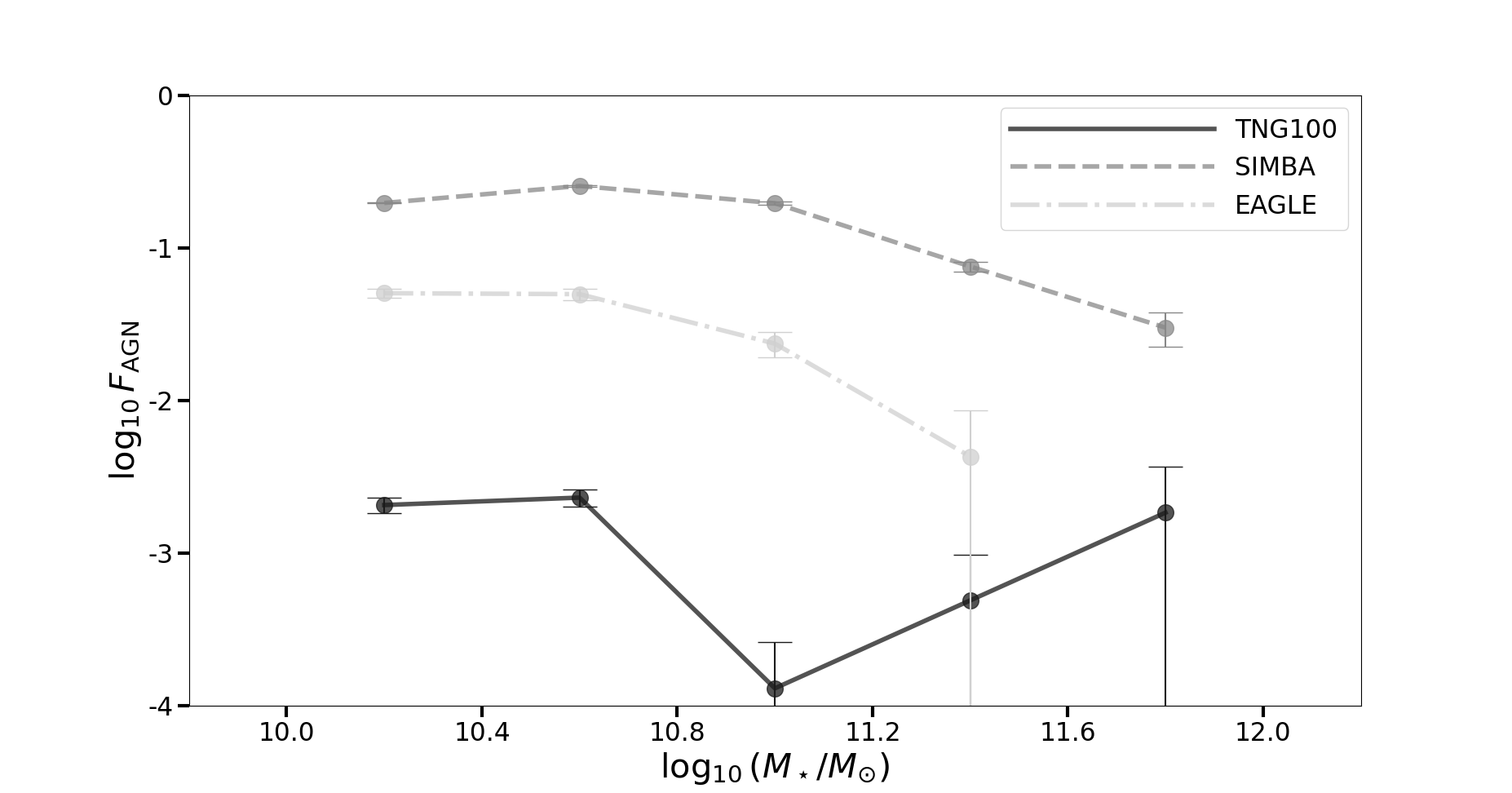}
    \caption{Comparison of the predicted $F_{\rm AGN}(M_\star)$ trends at $\lambda_{\rm c} = 10^{-3}$ for the EAGLE, SIMBA, and TNG100 simulations. The dash-dot line represents EAGLE, the dashed line corresponds to SIMBA, and the solid line denotes TNG100. Error bars indicate the $1\sigma$ uncertainty around the mean, assuming a binomial distribution. This figure demonstrates the ability of $F_{\rm AGN}$ to effectively distinguish between the AGN feedback implementations adopted in the different simulations.
}
    \label{fig:Fagn_models}
\end{figure*}

Additionally, it is important to emphasize that the simulations exhibit qualitatively distinct trends, as illustrated in Figure~\ref{fig:Fagn_models}. This figure presents the predicted mean $F_{\rm AGN}(M_\star)$ at $\lambda_{\rm c} = 10^{-3}$ for the three simulation models. The dash-dot line corresponds to EAGLE, the dashed line to SIMBA, and the solid line to TNG100. Error bars indicate the $1\sigma$ spread around the mean, assuming a binomial distribution. From the figure, it is evident that EAGLE consistently predicts higher values of $F_{\rm AGN}$ than TNG100 across all stellar masses, with an RMS deviation of 1.56 dex. SIMBA predicts even higher $F_{\rm AGN}$ values than both EAGLE and TNG100 at all masses, with an RMS deviation of 2.21 dex relative to TNG100. These discrepancies underscore the sensitivity of $F_{\rm AGN}$ to differences in AGN feedback prescriptions, highlighting its utility in differentiating between the feedback models employed by various simulations.

\subsection{Modified TNG100 Model}\label{ssec:modified_tng100}

The results of the previous section compel us to question the validity 
of the underlying physics of these black hole feedback models. Further, 
they also suggest that a careful reassessment of the current black hole 
feedback models may be necessary, either at the level of the specifics 
of the model or at a more fundamental level, concerning the basic 
underlying assumptions themselves. 

As a preliminary step along this direction, in this section we investigate 
the implications of modifying the specific details of one of the models. 
That is, we investigate whether, by tweaking its parameters, a given 
model can predict $F_{\rm AGN}$ trends that are in line with the observations,
given the galaxy and black hole properties of the simulation at a given
time step. If so, then further work would need to be done to understand
whether such a change could work self-consistently within the simulations 
while still reproducing the observational constraints originally used 
to tune the simulation parameters. 

Here we will perform the first step of this exercise
with the TNG100 black hole feedback model. Testing further with self-consistent
simulations is a much larger project beyond the scope of this work.

The black hole feedback model of TNG100 is explained in detail in Section \ref{sssec:tng100_bh_model}. Radio mode feedback for a black hole is active when $\dot{M} < \chi$ (see Equation \ref{eq:default_transition_chi}). To modify the radio mode feedback and have it match our $F_{\rm AGN}$ constraint, we modify the transition $\chi$ to $\chi_{\rm Q}$ for quiescent galaxies and $\chi_{\rm SF}$ for star forming galaxies:

\begin{equation}\label{eq:modified_transition_chi}
\chi_{\rm Q}, \chi_{\rm SF} = \min\left( \chi_{0} \left( \frac{M_{\rm BH}}{M_{0}} \right)^{\beta}, \chi_{\rm max} \right)
\end{equation}

We will change the parameters $\theta = ~$\{$\chi_{\rm 0}$, $M_{\rm 0}$, $\beta$, $\chi_{\rm max}$, $\Delta \log_{\rm 10} \left( \lambda_{\rm c} \right)$ \} from their default values of \{$0.002$, $10^{8}~M_{\odot}$, 2, $0.2$, 0\}. $\Delta \log_{\rm 10} \left( \lambda_{\rm c} \right)$ is not an intrinsic parameter of the feedback model but has been included here to allow for the potential variation in absolute scales of activity between observations and simulations (see Section \ref{ssec:d&m_overview}). 

Additionally, we will distinguish between quiescent and star-forming galaxies in our treatment. While this distinction is not driven by a specific physical motivation, it provides the model with the best opportunity to fit the data effectively. Specifically, we select two new sets of values for $\theta$—one for quiescent galaxies and one for star-forming galaxies---chosen to yield better agreement with our $F_{\rm AGN} \left(M_{\star} \right)$ constraints. In particular, we independently select two sets of values for $\theta$ such that they minimize a $\chi^2$ statistic  
quantifying the discrepancy between the observed and model-predicted values of $F_{\rm AGN}$ (and not to be confused with the default transition threshold parameter $\chi$). We define this $\chi^2$ as:
\begin{equation}\label{eq:chi2_stat}
\chi^2 = \sum_{i =1}^{n} \frac{\left( N_{\mathrm{AGN},~ i}^{\prime}(\theta) - N_{\mathrm{AGN},~ i}\right)^2} {\sigma_i^2}
\end{equation}
Here, $i$ denotes the $i^{\rm th}$ galaxy bin  with $N$ galaxies and there are $n = 4$ bins with detected AGN for star forming galaxies and $n = 3$ bins with detected AGN for quiescent galaxies. That is, we again only use those bins in which we have $1 \sigma$ confidence intervals on $F_{\rm AGN}$ and omit the bins in which all we have are upper limits on $F_{\rm AGN}$. For a given $\theta$, $N_{\mathrm{AGN},~ i}^{\prime}(\theta)$ and $F_{\mathrm{AGN},~ i}^{\prime}(\theta)$ are the model predicted number and fraction of AGN with $ \log_{\rm 10} \left(\lambda_{\rm } \right) > -3 + \Delta \log_{\rm 10} \left(\lambda_{\rm c} \right)$. The two are related by $N_{\mathrm{AGN},~ i}^{\prime}(\theta) = N \times F_{\mathrm{AGN},~ i}^{\prime}(\theta)$. $N_{\mathrm{AGN},~ i}$ is the number of AGN with $\log_{\rm 10} \left( \lambda \right) > -3$ in our TNG100 sample as predicted by the median $F_{\rm AGN} \left( M_\star\right)$ observational constraint, and, $N_{\mathrm{AGN},~ i} = N \times F_{\mathrm{AGN},~ i}$. Finally $\sigma_i$ is the binomial uncertainty on $N_{\mathrm{AGN},~ i}$. 

To minimize $\chi^2$ over $\theta$, we make use of the {\tt minimize} function (running the Nelder Mead algorithm) in python's {\tt scipy.optimize} module. We find that the minima occurs at $\theta_{\rm SF} = $ \{$\chi_{\rm 0}=0.09$, $M_{\rm 0} = 10^{10.48}$, $\beta = 1.95$, $\chi_{\rm max} = 0.0008$, $\Delta \log_{\rm 10} \left( \lambda_{\rm c} \right) = -0.89$\} for star forming galaxies and at $\theta_{\rm Q} = $ \{$\chi_{\rm 0}=0.08$, $M_{\rm 0} = 10^{10.12}$, $\beta = 2.00$, $\chi_{\rm max} = 0.001$, $\Delta \log_{\rm 10} \left( \lambda_{\rm c} \right) = -0.93$\} for quiescent galaxies. Even though the objective function is not particularly pleasant, the optimizations seems to be fairly stable and insensitive to the initial guess for $\theta$ over a broad range of the parameter space.

Ideally it would be best to re-run the simulation with the modified model to learn about the implications of the change. However, we will not run the simulation again in this study. Rather, we will make educated guesses about how the predictions of the simulation might change if everything about it stayed the same except the AGN feedback model. 

\begin{figure*}[htbp]
  \gridline{
    \fig{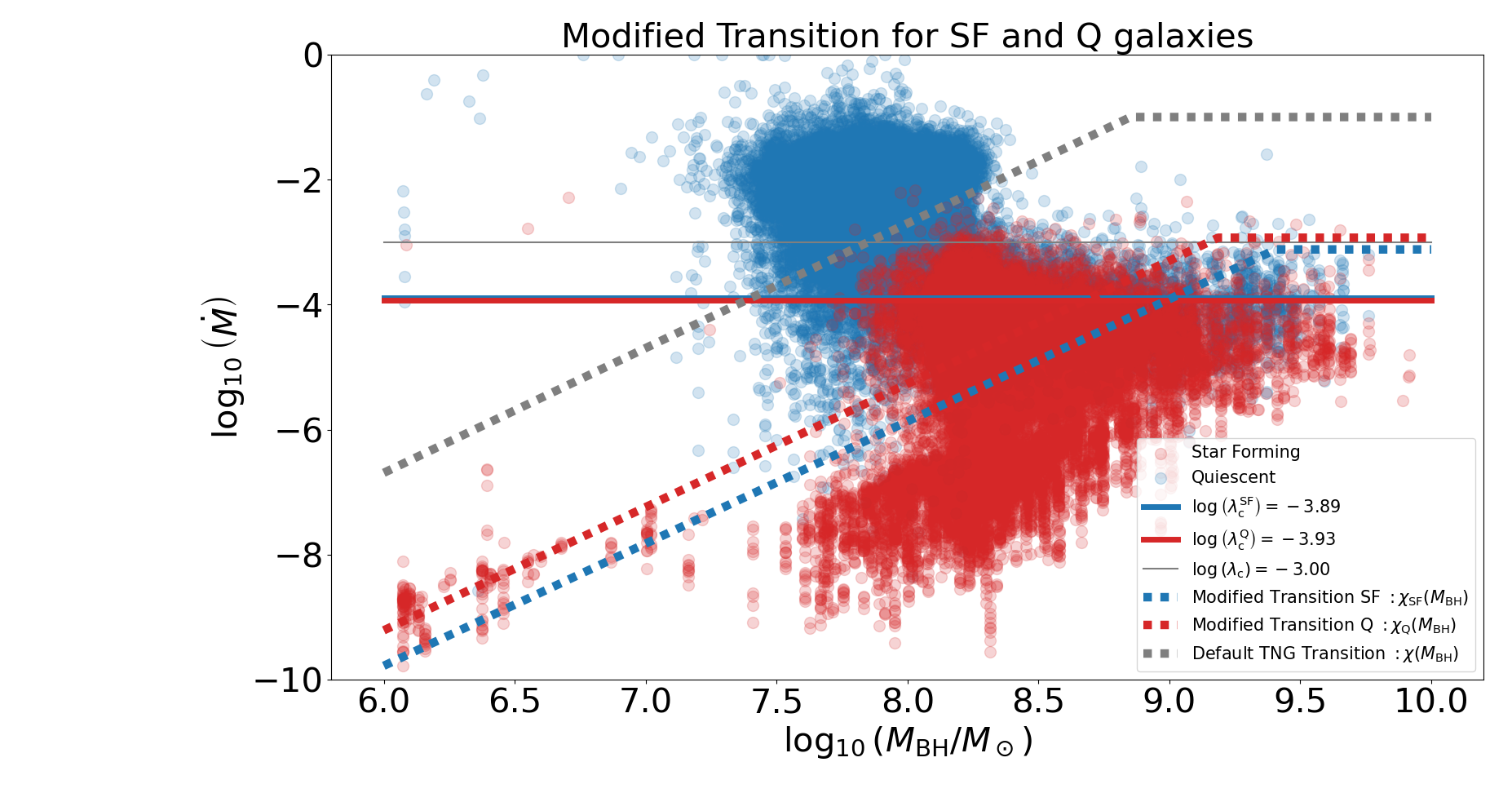}{0.95\textwidth}{
      $\dot{M}$ versus $M_{\rm BH}$ for our TNG100 sample. The thick gray dashed line represents the default TNG100 transition curve, $\chi(M_{\rm BH})$, while the thin gray solid line corresponds to $\log_{10}(\lambda_{\rm c}) = -3$. The thick red dashed line indicates the modified transition curve for quiescent galaxies, $\chi_{\rm Q}(M_{\rm BH})$, with the associated thin red solid line representing $\log_{10}(\lambda^{\rm Q}_{\rm c}) = -3.93$. Similarly, the thick blue dashed line shows the modified transition curve for star-forming galaxies, $\chi_{\rm SF}(M_{\rm BH})$, and the thin blue solid line corresponds to $\log_{10}(\lambda^{\rm SF}_{\rm c}) = -3.89$. For a given model, galaxies above the transition line are in the quasar mode of AGN feedback, while those below the line are in the radio mode. Galaxies located below the default transition line but above the modified would host radio-mode AGN under the default model, but not under the modified prescriptions.
    }
  }
  \gridline{
    \fig{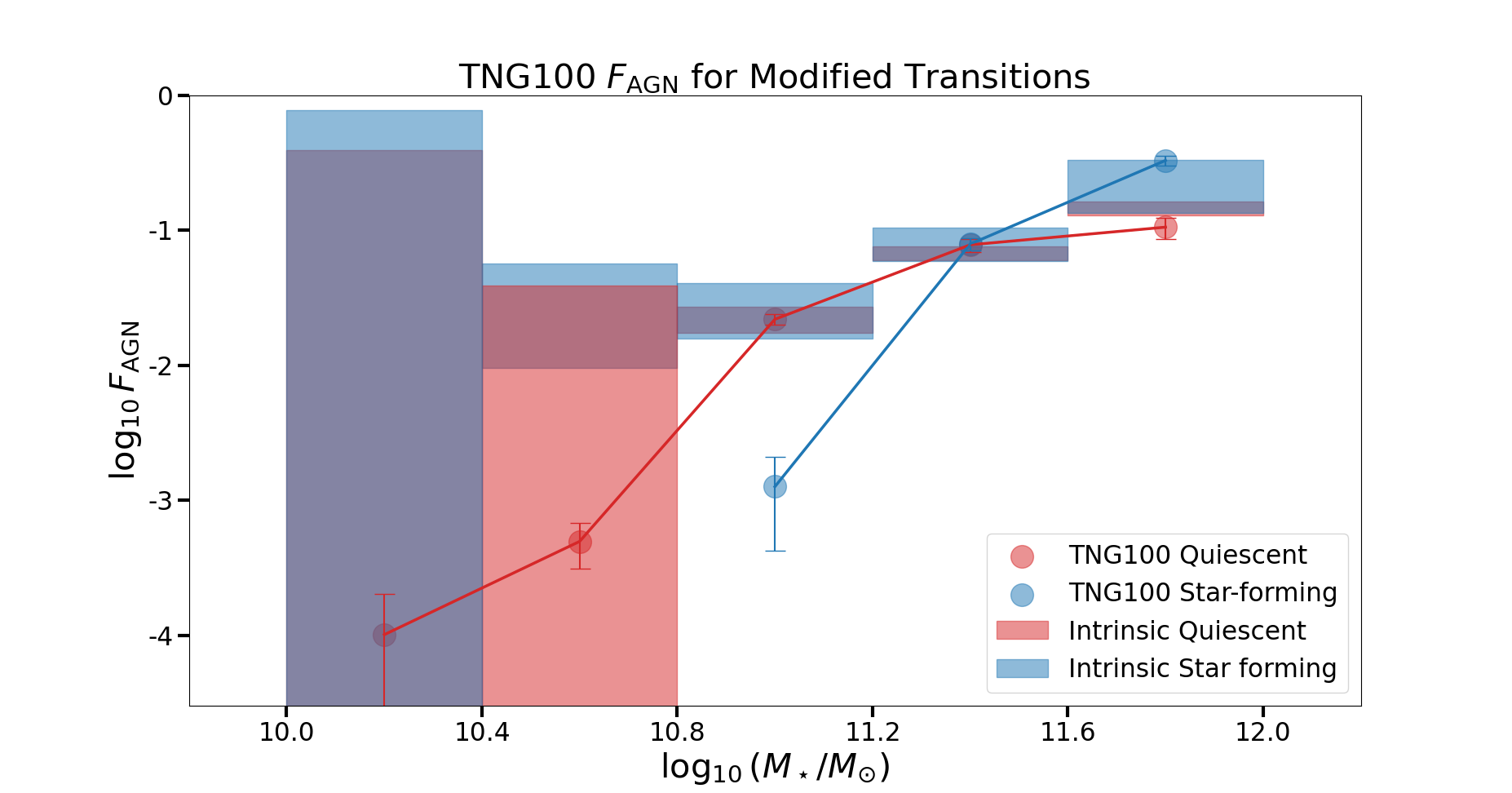}{0.95\textwidth}{
      Predictions for $F_{\rm AGN}(M_\star)$ under the modified models for quiescent and star-forming galaxies compared to our observational constraint shown in Figure \ref{fig:Fagn_mbins}. The predictions now show a strong dependence on $M_\star$. However, these modifications are likely to impact the key observational constraints to which the original model was calibrated (see Section~\ref{ssec:modified_tng100} and the top panel of this figure).
    }
  }
  \caption{Modified TNG100 black hole feedback transition lines for star-forming and quiescent galaxies (top), and the corresponding predictions for $F_{\rm AGN} \left(M_\star \right)$ (bottom).}
  \label{fig:modified_tng100}
\end{figure*}

Figure \ref{fig:modified_tng100} shows the results of the $\chi^2$ optimization. The top panel shows $\dot{M}$ plotted against $M_{\rm BH}$ for all the galaxies in our TNG100 sample. The blue dots are star forming galaxies and the red dots are quiescent galaxies. The gray dashed line corresponds to the default transition $\chi$ of the TNG100 model and the thin gray solid line denotes $\log_{\rm 10} \left(\lambda_{\rm c} \right) = -3$\footnote{The y-axis is an Eddington scaled mass accretion rate whereas $\lambda_{\rm c}$ is an Eddington ratio. For a given black hole, the two types of quantities are related by $\lambda = \epsilon_{\rm kin}\dot{M}/\epsilon_{\rm r}$. $\epsilon_{\rm kin}$ can depend on the gas density around the black hole. However, for almost all the galaxies in this TNG100 sample $\epsilon_{\rm kin} \sim 0.2$ and hence $\dot{M} \sim \lambda$. We have made this approximation in here, especially while showing $\log_{\rm 10} \left(\lambda_{\rm c} \right)$, $\log_{\rm 10} \left(\lambda^{\rm Q}_{\rm c} \right)$ and $\log_{\rm 10} \left(\lambda^{\rm SF}_{\rm c} \right)$.}. The thick blue dashed line denotes the modified transition $\chi_{\rm SF}$ for star forming galaxies and thin blue solid line corresponds to $\Delta \log_{\rm10} \left(\lambda^{\rm SF}_{\rm c}\right) = -0.89$. The thick red dashed line denotes the modified transition $\chi_{\rm Q}$ for quiescent galaxies and thin red solid line corresponds to $\Delta \log_{\rm10} \left(\lambda^{\rm Q}_{\rm c}\right) = -0.93$.

The bottom panel shows the predicted $F_{\rm AGN}$ values for the modified TNG100 models. The modified models clearly succeed in capturing the strong dependence of $F_{\rm AGN}$ on $M_\star$ for both star-forming and quiescent galaxies. The most significant improvement is in the slope of the predicted $F_{\rm AGN}$ for quiescent galaxies, which now has a value of approximately 0.85. Although the modified model still captures the strong $M_\star$ dependence for star-forming galaxies, the resulting slope is now $\sim 3.02$—steeper than what is observed. As a consequence, the model's performance with respect to the sSFR independence has slightly degraded, with the RMS deviation between the red and blue curves increasing from 0.32 to 0.38. 

Additionally, the modified model is likely to produce a galaxy stellar mass function and star formation rate density that deviate from those the default model was originally calibrated to match. This is clearly illustrated in the top panel of Figure \ref{fig:modified_tng100}, where a significant fraction of quenched galaxies ($\sim 35 \%$) and star forming galaxies ($\sim 25 \%$) would no longer host radio AGN after the model is modified (indicated by red circular points located above $\chi_{\rm Q}$ and below $\chi$, and the blue circular points above $\chi_{\rm SF}$ and below $\chi$). 
For the majority of the TNG100 galaxies studied here (\( \sim 90\% \)), the ratio \( \epsilon_{\rm kin}/\epsilon_{\rm r} \epsilon_{\rm f} \) is approximately 10. In other words, at fixed \( \dot{M} \), the radio-mode feedback energy couples significantly more efficiently to the host galaxy compared to the quasar mode. As a result, radio-mode feedback is essential in TNG100 for quenching galaxies. Under the modified model described here, a substantial fraction of galaxies would have relatively higher star formation rates as compared to the default model. As a result, a significant fraction of galaxies that were originally quenched would no longer remain so, leading to changes in key observables used to calibrate TNG100, specifically the star formation rate density and the stellar mass function at \( z = 0 \).
In other words, the chosen form of the TNG100 black hole feedback model is likely not suitable for simultaneously reproducing the observed stellar mass function, star formation rate density, and the $F_{\rm AGN}$ trend. A reassessment of the fundamental assumptions behind the TNG100 black hole feedback model may therefore be necessary.

\section{Conclusion}\label{sec:conclusion}

We have tested the black hole feedback models of the EAGLE, SIMBA and TNG100 cosmological simulations to evaluate their underlying physics by studying their radio AGN fractions. Specifically, we use the fraction of galaxies $F_{\rm AGN}$ that are radio AGN with an Eddington ratio $\lambda > \lambda_{c}$, where $\lambda_c = 10^{-3}$, as a function of $M_{\star}$ (see Sections \ref{intro} and \ref{sec:Fagn_constraint}; \citealt{sureshblanton2024}). On comparison, we find that:

\begin{itemize}
    \item None of the three models, for any value of $\lambda_{c}^{\prime}$, predict $F_{\rm AGN}$ trends that are in line with the observations. In particular, the model predicted $F_{\rm AGN}$ curves fail to recreate the strong observed dependence on $M_{\star}$ and lack thereof on sSFR (see Sections \ref{sec:d&m} and \ref{sec:results}). 

    \item The simulations almost always predict $F_{\rm AGN}\left(M_{\star}\right)$ curves that have slopes $\lesssim$ 0, in stark contrast to the observed positive slope $\sim 1.03$ for quiescent galaxies and $\sim 0.85$ for star forming galaxies.

    \item Figure~\ref{fig:Fagn_models} highlights the utility of the $F_{\rm AGN}$ metric in distinguishing between various AGN feedback models. EAGLE consistently predicts higher values of $F_{\rm AGN}$ than TNG100 across the full range of stellar masses, with an RMS deviation of 1.56~dex. Furthermore, SIMBA predicts systematically higher $F_{\rm AGN}$ values across the entire stellar mass range compared to both EAGLE and TNG100, with an RMS deviation of 2.21~dex relative to TNG100.

    \item The given form of the TNG100 model, if specifically optimized to reproduce the observed strong mass dependence of $F_{\rm AGN}$, is able to do so. However, this most likely will alter the predicted galaxy stellar mass function and star formation rate density---key observational constraints the model was calibrated to in the first place (see Section \ref{ssec:modified_tng100}). This likely means that the TNG100 class of models is inadequate to simultaneously predict the three observational constraints mentioned above. Hence, a reassessment of the fundamental assumptions in the TNG100 model may be necessary. 
    
\end{itemize}

The simulations generally either overestimate or underestimate the amount of radio-mode feedback relative to observational data, as shown in Figures~\ref{fig:Fagn_eagle}, \ref{fig:Fagn_simba}, and \ref{fig:Fagn_illustris}. This discrepancy may arise because the simulated galaxies possess nuclear gas densities that differ from those typically observed in real galaxies, or because similar gas densities do not trigger radio AGN activity in the simulations as they do in reality. A promising direction for future research is to directly compare the link between AGN activity and gas densities on $\sim$100 pc scales in both simulations and observations.

Furthermore, in this study we have ignored the quasar mode AGN population and its relationship to host properties. In simulations, quasar mode feedback at earlier cosmic times typically acts to expel gas from the central regions of galaxies. If this feedback is too weak (or too strong), it may lead to a surplus (or deficit) of gas in the galactic center, resulting in central gas densities that are higher (or lower) than observed in real galaxies.
To evaluate the quasar-mode feedback implemented in simulations, studies similar to \cite{sureshblanton2024}---as well as the work presented here---but focused on narrow-line or X-ray AGN populations, would be highly valuable. Combined with our current analysis, such efforts would facilitate a more comprehensive assessment of contemporary black hole feedback models.

\section*{Acknowledgements}

We thank the EAGLE, SIMBA, and IllustrisTNG teams for making their data public. We thank David Hogg, Alexander Novara, Valentina Tardugno, Connor Hainje, and Matt Daunt of New York University. We thank Sophie Koudmani of Cambridge University and Beatriz Mingo of University of Hertfordshire for their time and valuable feedback about this project. We also thank the Galaxies group at the Center for Computational Astrophysics and the Black Hole Mapper AGN Population group of SDSS-V for their valuable insights that have helped shape this study.

EAGLE used the DiRAC Data Centric system at Durham University, operated by the Institute for Computational Cosmology on behalf of the STFC DiRAC HPC Facility (\url{www.dirac.ac.uk}); this equipment was funded by BIS National E-infrastructure capital grant ST/K00042X/1, STFC capital grant ST/H008519/1, STFC DiRAC Operations grant ST/K003267/1 and Durham University. DiRAC is part of the National E-Infrastructure. The study was sponsored by the Dutch National Computing Facilities Foundation (NCF) for the use of supercomputer facilities, with financial support from the Netherlands Organisation for Scientific Research (NWO), and the European Research Council under the European Union’s Seventh Framework Programme (FP7/2007–2013) / ERC Grant agreements 278594 GasAroundGalaxies, GA 267291 Cosmiway, and 321334 dustygal. Support was also received via the Interuniversity Attraction Poles Programme initiated by the Belgian Science Policy Office ([AP P7/08 CHARM]), the National Science Foundation under Grant No. NSF PHY11-25915, and the UK Science and Technology Facilities Council (grant numbers ST/F001166/1 and ST/I000976/1) via rolling and consolidating grants awarded to the ICC.

The SIMBA simulation was run on the DiRAC@Durham facility managed by the Institute for Computational Cosmology on behalf of the STFC DiRAC HPC Facility. The equipment was funded by BEIS capital funding via STFC capital grant nos ST/P002293/1, ST/R002371/1, and ST/S002502/1, Durham University, and STFC operations grant no. ST/R000832/1. DiRAC is part of the National e-Infrastructure.

The IllustrisTNG simulations were undertaken with compute time awarded by the Gauss Centre for Supercomputing (GCS) under GCS Large-Scale Projects GCS-ILLU and GCS-DWAR on the GCS share of the supercomputer Hazel Hen at the High Performance Computing Center Stuttgart (HLRS), as well as on the machines of the Max Planck Computing and Data Facility (MPCDF) in Garching, Germany.

Funding for the Sloan Digital Sky Survey IV has been provided by the Alfred P. Sloan Foundation, the U.S. Department of Energy Office of Science, and the Participating Institutions.

SDSS-IV acknowledges support and resources from the Center for High Performance Computing at the University of Utah. The SDSS website is \url{www.sdss4.org}.

SDSS-IV is managed by the Astrophysical Research Consortium for the Participating Institutions of the SDSS Collaboration including the Brazilian Participation Group, the Carnegie Institution for Science, Carnegie Mellon University, Center for Astrophysics | Harvard \& Smithsonian, the Chilean Participation Group, the French Participation Group, Instituto de Astrof\'isica de Canarias, The Johns Hopkins University, Kavli Institute for the Physics and Mathematics of the Universe (IPMU) / University of Tokyo, the Korean Participation Group, Lawrence Berkeley National Laboratory, Leibniz Institut f\"ur Astrophysik Potsdam (AIP), Max-Planck-Institut f\"ur Astronomie (MPIA Heidelberg), Max-Planck-Institut f\"ur Astrophysik (MPA Garching), Max-Planck-Institut f\"ur Extraterrestrische Physik (MPE), National Astronomical Observatories of China, New Mexico State University, New York University, University of Notre Dame, Observat\'orio Nacional / MCTI, The Ohio State University, Pennsylvania State University, Shanghai Astronomical Observatory, United Kingdom Participation Group, Universidad Nacional Aut\'onoma de M\'exico, University of Arizona, University of Colorado Boulder, University of Oxford, University of Portsmouth, University of Utah, University of Virginia, University of Washington, University of Wisconsin, Vanderbilt University, and Yale University.

\bibliography{refs.bib}

\end{document}